\documentclass[aps,prx,reprint,superscriptaddress]{revtex4-2}
\usepackage[T1]{fontenc} 
\usepackage{microtype, fancyhdr}
\usepackage[utf8]{inputenc} 
\usepackage[english]{babel}
\usepackage{siunitx}
\usepackage{amsmath}
\usepackage{lmodern}
\usepackage{graphicx, amsmath, wrapfig, subcaption}
\usepackage{chngcntr}
\usepackage{amssymb, enumitem, hhline}
\usepackage{braket, array}
\usepackage{multirow}
\usepackage[version=4]{mhchem}
\usepackage{makecell}
\usepackage{ragged2e}
\makeatletter
\long\def\@makecaption#1#2{%
  \vskip\abovecaptionskip
  \small
  \noindent\textbf{#1. }\justifying #2\par
  \vskip\belowcaptionskip
}
\makeatother
\newcommand{\ketbra}[2]{\ket{#1}\bra{#2}}

\begingroup\lccode`~=`!
\lowercase{\endgroup\def~}#1{_{\mathrm{#1}}}
\mathchardef\exclam=\mathcode`!
\AtBeginDocument{\mathcode`!=\string"8000 }

\usepackage{xcolor} % Enable \definecolor
\definecolor{NewBlue}{rgb}{0, 0, 0.41}
\definecolor{NewRed}{rgb}{0.6, 0.07, 0.07}
\usepackage[colorlinks,
    linkcolor=NewBlue,
    citecolor=NewBlue,
    urlcolor=NewRed]{hyperref}
\usepackage{orcidlink}
    
\begin{document}
\title{Remote Entanglement of Solid-State Spin Qubits Integrated in Broadband Waveguides
%A quantum network link based on waveguide-integrated solid-state qubits
}
\newcommand{\EqualContrib}{These authors contributed equally to this work}
\author{Christopher~\surname{Waas }\orcidlink{0009-0008-1878-2051}}
\thanks{\EqualContrib}
\author{Timo~\surname{Dolné}\orcidlink{0009-0000-3022-7351}}
\thanks{\EqualContrib}
\author{Hans~K.~C.~\surname{Beukers}\orcidlink{0000-0001-9934-1099}}
\thanks{\EqualContrib}
\author{Alexander~M.~\surname{Stramma}\orcidlink{0009-0001-5400-9575}}
\thanks{\EqualContrib}
\author{Nina~\surname{Codreanu}\orcidlink{0009-0006-6646-8396}}
\author{Noé~\surname{Mathieu}\orcidlink{0009-0008-3425-3088}}
\author{Ronald~\surname{Hanson}\orcidlink{0000-0001-8938-2137}}
\email{R.Hanson@TUDelft.nl}
 
\affiliation{QuTech and Kavli Institute of Nanoscience, Delft University of Technology, P.O. Box 5046, 2600 GA Delft, The Netherlands}
% \date{2026}

\begin{abstract}
\noindent
Solid-state spin-photon interfaces promise to scale quantum networks through on‑chip photonic integration and multiplexed entanglement generation. To date, remote entanglement between integrated emitters has been realized only in cavity‑enhanced systems, where fabrication yield and spectral matching remain major obstacles. Here we demonstrate heralded remote entanglement between diamond tin-vacancy spin qubits embedded in separate on-chip waveguides. Combining intrinsically efficient photon emission with a broadband waveguide architecture provides high device yield and obviates the need for spectral matching to cavity modes. We realize coherent optical and spin control and achieve high-visibility two-photon interference. By combining photon-mediated entanglement generation with real-time feedforward, we produce a consistent entangled state independent of the heralding pattern. These results establish waveguide-integrated tin-vacancy centers as a compelling platform for scalable quantum network nodes.
\end{abstract}

\maketitle

\section{Introduction}
Quantum networks that distribute entanglement between distant processing nodes promise applications in secure communication, low-latency coordination, distributed computation, and quantum-enhanced sensing \cite{kimbleQuantumInternet2008a, wehnerQuantumInternet2018,
komarQuantumNetwork2014}.
Recent demonstrations across multiple platforms have realized key networking primitives in the laboratory \cite{daissQuantumlogicGateDistant2021, zhangDeviceindependentQuantumKey2022a, mainDistributedQuantumComputing2025, liuLonglivedRemote2026, stasEntanglementassistedNonlocal2026, iulianoUnconditionallyTeleported2026a} 
and have distributed entanglement between qubit nodes separated by metropolitan distances using deployed telecom fiber \cite{liuCreationMemory2024, stolkMetropolitanscaleHeralded2024}. Scaling these capabilities to real-world applications will require high‑density multiplexing of remote entanglement generation to achieve practical entanglement distribution rates.

On-chip photonic integration of solid-state emitters - combining small footprint with semiconductor fabrication techniques - provides a compelling route. In this direction, remote spin-spin entanglement has recently been achieved with integrated solid‑state emitters in resonant nanophotonic cavities \cite{knautEntanglementNanophotonic2024, ruskucMultiplexedEntanglement2025, photonicinc2024distributed}; however, limited device yield and the need for precise cavity-resonance tuning remain significant hurdles to scaling.

\begin{figure}
    %\centering
    \includegraphics[width=\columnwidth]{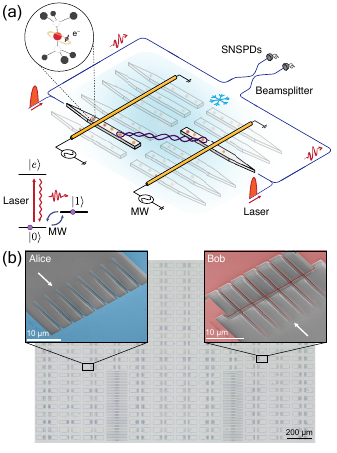}
    \caption{\textbf{Quantum network platform.} \textbf{(a)} Schematic of the platform. Diamond tin-vacancy spin qubits, with inversion symmetric structure (upper inset), are integrated into broadband diamond waveguide devices, operated around $T=\qty{1}{\K}$ in a magnetic field of \qty{117}{mT} (Appendix \ref{app: magnetic field}). The microwave (MW) and laser fields for spin qubit and optical control (lower inset) are routed to individual devices via separate wirebonds and tapered fibers, respectively. Entanglement is generated by interference of spin-entangled photons at a central beamsplitter and detection via superconducting nanowire single-photon detectors (SNSPDs). \textbf{(b)} Optical microscope image of the diamond chip. Unit cells, each containing multiple waveguides, are assorted in columns. Inset: False-colored scanning electron microscope images of the waveguides used for the results in the main text.}
    \label{fig:device}
\end{figure}

Here we demonstrate a quantum network link based on diamond tin‑vacancy (SnV) centers integrated into broadband waveguides (Fig.~\ref{fig:device}(a)). Leveraging the favorable optical and spin properties of SnV centers \cite{trusheimTransformLimitedPhotons2020, gorlitzSpectroscopicInvestigations2020, arjonamartinezPhotonicIndistinguishability2022, pasiniNonlinearQuantum2024, guoMicrowaveBasedQuantum2023, rosenthalMicrowaveSpin2023, karapatzakisMicrowaveControl2024}, this architecture supports high fabrication yield and removes the need for active cavity tuning. We realize independent coherent optical and spin control on two SnV centers in separate waveguides and apply resonant optical excitation and collection to demonstrate two‑photon quantum interference. Using this complete toolbox, we generate heralded remote spin-spin entanglement through spin-photon entanglement followed by photon interference and detection. Together with recently demonstrated approaches for photonic integration \cite{wanLargescaleIntegration2020,liHeterogeneousIntegration2024}, on‑chip frequency tuning \cite{clarkNanoelectromechanicalControl2024,brevoordLargerangeTuning2025} and quantum frequency conversion to the telecom band \cite{brevoordQuantumFrequency2025}, our results unlock a viable path to highly multiplexed quantum network links.

\section{Integrated quantum network nodes}
\begin{figure*}
    \centering
    \includegraphics{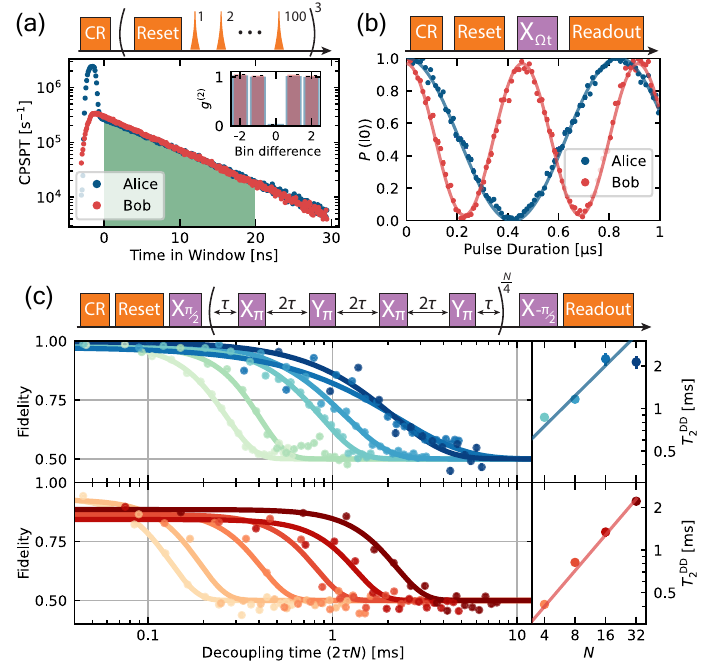}
    \caption{\textbf{Characterization of the quantum network node.} Pulse sequences used are shown above each panel. Each sequence starts with a charge-resonance (CR) check and spin reset pulse. \textbf{(a)} Fast optical excitation and single-photon detection.
    Photon detection probability density, measured in counts per shot per time (CPSPT), as a function of time. The photon detection window, starting right after the optical $\pi$-pulse, is indicated in green. Inset: second-order auto-correlation functions $g^2$ showing high photon purity.
    \textbf{(b)} Spin qubit control. An applied MW-drive induces Rabi oscillations.
    \textbf{(c)} Spin qubit coherence. After preparing the qubit in a balanced superposition,  XY decoupling sequences are applied with varying inter-pulse delays and pulse numbers. Left: fidelity to initial superposition as function of total decoupling time. 
    Right: scaling of spin coherence time $T_2$ with number of pulses $N$. Upper (lower) panels correspond to Alice (Bob).}
    \label{fig:characterization}
\end{figure*}
Our quantum network nodes are based on the integration of SnV centers in diamond single-mode waveguides. The waveguides are fabricated on a high-purity diamond chip using lithography and etching techniques \cite{khanalilooHighQMonolithic2015}. Crucially, the broadband nature of the waveguide modes yields large fabrication tolerance that enables a yield of virtually unity. Here, we fabricate a chip containing 8,336 waveguides with a structural yield of $\qty{99}{\%}$ (Fig.~\ref{fig:device}(b)). In the experiments presented we couple the waveguide modes directly to tapered optical fibers in lensed mode. Similar waveguides can be hybrid integrated into photonic chips by transfer printing \cite{wanLargescaleIntegration2020} or by fabricating directly on chip-bonded diamond membranes \cite{guoDirectbondedDiamond2024, riedelScalablePhotonic2026}, providing a clear path for further on-chip scaling and integration.

In our device, SnV centers were created in the diamond by ion implantation and annealing prior to waveguide fabrication. At cryogenic temperature, the negatively charged SnV center (used throughout this work) forms an effective spin-1/2 system (defining the qubit states) with spin-conserving optical transitions at a wavelength of  $\qty{619}{\nm}$. Owing to its inversion symmetry, the SnV center is insensitive to electric charge noise at first order, allowing for on-chip integration in close proximity to surfaces without significant degradation of optical coherence \cite{trusheimTransformLimitedPhotons2020, arjonamartinezPhotonicIndistinguishability2022,pasiniNonlinearQuantum2024}. Furthermore, the SnV center has an intrinsic coherent photon emission probability of $\approx 0.36$ \cite{iwasakiTinVacancyQuantum2017, gorlitzSpectroscopicInvestigations2020}, among the highest of the well-studied color center qubits \cite{dohertyNitrogenvacancyColour2013, neuPhotophysicsSingle2012, beckerCoherenceProperties2017, kindemControlSingleshotReadout2020}. Combined with a relatively fast optical decay (lifetime $\approx \qty{7}{\ns}$), this provides an efficient optical interface without relying on resonant cavity structures. The 52 waveguides investigated in this study contain about one optically coherent (linewidth $<\qty{50}{\MHz}$) and well-coupled SnV center on average. We choose two SnV centers from opposite sides of the chip, hereafter labeled A(lice) and B(ob), for the main results below.

The SnV optical transitions enable spin initialization (by optical pumping) and single-shot readout (through state-dependent fluorescence) as well as single-photon generation for establishing remote entanglement. We implement spin-selective optical $\pi$-pulses using short resonant laser pulses (Appendix~\ref{app:pulses}). For detecting the resulting spontaneously emitted photon while rejecting reflected laser light, we employ a combination of cross-polarization and detection time filtering. We show resulting photon detection time histograms for Alice and Bob in Fig.~\ref{fig:characterization}(a). We define a $\qty{20}{\ns}$ detection window after the pulse that is used for all the results below. The detected photonic modes exhibit high purity as evidenced by a second-order auto-correlation function $g^2 (0)$ of $0.036(2)$ $(0.0042(6))$ for Alice (Bob) (inset of Fig.~\ref{fig:characterization}(a)). The remaining impurity is attributed to off-resonant excitation of other emitters in the waveguide (for Alice) and to imperfect laser rejection. These could be mitigated in future devices by reducing the number of emitters using masked implantation, and by improving the coupling efficiency to the waveguide using for instance in-contact coupling \cite{burekFiberCoupledDiamond2017}.

The spin qubit states are coherently controlled using resonant microwave pulses (Fig.~\ref{fig:characterization}(b)) \cite{rosenthalMicrowaveSpin2023, guoMicrowaveBasedQuantum2023}. We probe the coherence of the qubits under dynamical decoupling (Fig.~\ref{fig:characterization}(c)) and find, similar to previous work \cite{rosenthalMicrowaveSpin2023, beukersControlSolidState2025, karapatzakisMicrowaveControl2024, strammaTinvacancyCentre2024}, that the coherence time $T_{2}$ scales with the number of decoupling pulses $N$ as $T_2^{DD} \propto N^\chi$. This scaling is typical for slowly varying spin bath environments, with the local spin distributions determining the precise scaling factors $\chi!A = \num{0.46(7)}$ and $\chi!B=\num{0.78(4)}$. Importantly, with energy relaxation only relevant at timescales beyond seconds \cite{trusheimTransformLimitedPhotons2020}, more advanced dynamical decoupling sequences can continue to improve coherence
\cite{karapatzakisMicrowaveControl2024, strammaTinvacancyCentre2024} while the initial coherence can be further increased through isotopic purification \cite{bradleyRobustQuantumnetwork2022}. In the current device, a single decoupling pulse already extends the coherence time ($T_2^\text{echo}>\qty{130}{\us}$) significantly beyond the duration of one entanglement generation attempt ($\approx \qty{17}{\us}$), making decoherence during entanglement generation negligible. For $N = 32$ pulses, both qubits exhibit a coherence time larger than $\qty{2}{\ms}$.
\begin{figure*}
    \centering
    \includegraphics{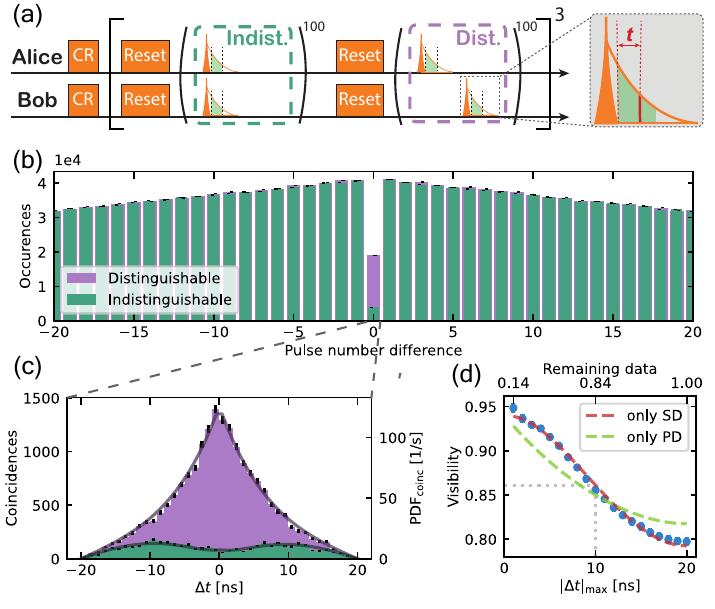}
    \caption{\textbf{Two-photon quantum interference.} \textbf{(a)} Pulse sequence. Indistinguishable (distinguishable) photon pairs are created by simultaneous (non-simultaneous) resonant optical excitation of the two SnVs.
    \textbf{(b)} Histogram of measured coincidences for the two scenarios as a function of excitation pulse difference. For a pulse difference of zero, indistinguishable photons interfere yielding a strong suppression of coincidences.
    \textbf{(c)} Histogram of measured coincidences for zero pulse difference as a function of detection time difference $\Delta t$. The solid lines are the coincidence probability density functions calculated from the fit in (d).
    \textbf{(d)} Visibility as a function of maximum detection time difference $|\Delta t|!{max}$. Limiting $|\Delta t|!{max}$ improves the visibility at the cost of data rate (top axis). Fits shown take into account, in addition to finite photon purity, only spectral diffusion (SD) or only pure dephasing (PD). Dotted line indicates value of $|\Delta t|!{max}$ used in the entanglement experiments.
    }
    \label{fig:tpqi}
\end{figure*}

\section{Two-photon quantum interference} \label{sec: TPQI}

Remote entanglement generation hinges on the interference of spin-entangled photons emitted by the two SnV centers, placing stringent requirements on the photon indistinguishability. In particular, for the two-photon Barrett-Kok protocol used here \cite{barrettEfficientHighfidelity2005}, the entangled state fidelity $\mathcal{F}$ is related to the two-photon interference visibility $V$ via $\mathcal{F} \leq (1+V)/2$, where the equality holds in absence of other error sources \cite{bernienHeraldedEntanglement2013}. Recent work has demonstrated interference of consecutively emitted photons from a single SnV center \cite{arjonamartinezPhotonicIndistinguishability2022} as well as two-photon interference using off-resonant (and thus not spin-selective) excitation \cite{bushmakinTwoPhotonInterference2024}, reporting interference visibilities up to \qty{80}{\%}. Here, we investigate two-photon interference under spin-selective resonant optical excitation as required for the entanglement generation protocol.

The indistinguishability of the photons is optimized by aligning all their degrees of freedom. For the main results presented below, we tune the emission frequencies of the SnV pair onto resonance by adapting the magnetic field strength and orientation (Appendix~\ref{app: magnetic field}). As an alternative, we also perform photon interference experiments by shifting the frequency of one of the emitted photons using electro-optical phase modulation, yielding consistent results (Appendix~\ref{app:AddTPQI}). While these methods suffice for aligning a single pair, recently demonstrated in-situ frequency tuning by electromechanical strain control \cite{clarkNanoelectromechanicalControl2024, brevoordLargerangeTuning2025}, compatible with our device geometry, enables practical frequency alignment of many emitters on-chip in a scalable and dynamical fashion.

Figure~\ref{fig:tpqi}(a) shows the experimental sequence. First, to ensure high optical coherence of each SnV center as well as precise optical frequency alignment between the two SnV centers, we apply charge-resonance checks. In this procedure, for each SnV, photon counts are recorded while simultaneously driving the two spin-cycling transitions and continuation is conditioned on surpassing a pre-defined threshold (Appendix~\ref{app:ple}). The sequence proceeds once both SnV centers have passed this check. Then, following a spin reset pulse, a series of optical $\pi$-pulses is applied to the two SnV centers. To maximize the temporal overlap of the emitted photons on the beamsplitter, and thus the indistinguishability, we optimize the relative timing of the pulses. Analogously, we measure the interference visibility of fully distinguishable photons - used for normalization - by shifting the excitation pulse on Bob in time such that the temporal overlap becomes negligible.
We label photon detections according to the corresponding number of the excitation pulse, ranging from 1 to 100. We then record the number of coincidence detections as a function of the difference of the two corresponding excitation pulse numbers. Only photons from the same excitation pulse (excitation pulse difference 0) can exhibit interference.

Figure~\ref{fig:tpqi}(b) shows the measured coincidences as a function of pulse number difference. The decrease in coincidences for higher absolute values of the difference is caused by the limited possible combinations in a block of $100$ excitation pulses. The visibility is given by $V = 1-(N_0^\text{indist}/N_0^\text{dist})$, where $N_0^\text{indist}$ ($N_0^\text{dist}$) is the number of coincidences with pulse number difference of 0 for the temporally overlapped (temporally separated) photons. We extract a raw visibility of $V=0.795(4)$.

\begin{figure*}
    \centering
    \includegraphics{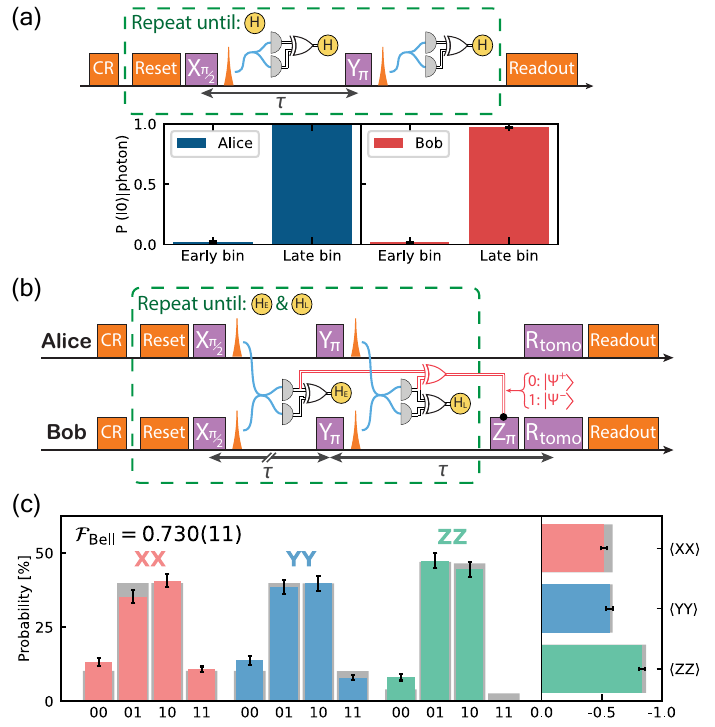}
    \caption{\textbf{Heralded remote spin-spin entanglement.} 
    \textbf{(a)} Spin-photon correlations. Top: Pulse sequence. Bottom: Qubit measurement outcomes split out per heralding optical pulse.
    \textbf{(b)} Heralded spin-spin entanglement pulse sequence. More details in Appendix~\ref{app:entanglement}.
    \textbf{(c)} Measured spin-spin correlations on heralded entangled states. Left: Measurement outcomes for the joint readout of the two spin qubits. Probabilities are corrected for known readout infidelities and based on 303, 259 and 260 heralding events in the XX, YY and ZZ bases. The grey bars behind the data represent the expected readout probabilities from simulation. Right: Calculated values of the correlators from the readout probabilities.
    }
    \label{fig:entanglement}
\end{figure*}
Common limitations to the visibility are imperfect photon purity (as discussed above), as well as spectral diffusion and pure dephasing of the optical transitions. The latter two effects can be mitigated by restricting the accepted time-difference $\Delta t$ between the photon detection events.
In Fig.~\ref{fig:tpqi}(c) we plot the coincidences of the zero-difference bin as a function of $\Delta t$, from which we extract the visibility as a function of maximum detection time difference $\left|\Delta t\right|!{max}$ (Fig.~\ref{fig:tpqi}(d)). We observe an increased visibility for smaller $\left|\Delta t\right|!{max}$, reaching $V = 0.948(5)$ at $\left|\Delta t\right|!{max}=\qty{1}{\ns}$. This value is on par with the highest reported for any solid-state emitter pair \cite{ruskucMultiplexedEntanglement2025}, but achieved here without a nanophotonic cavity. As no background corrections are applied here, the reported visibilities directly map to achievable entanglement fidelity. 

We fit the visibility data to a model that incorporates the effects of spectral diffusion and pure dephasing and includes the simultaneously measured photon purities (Appendix~\ref{app:model_tpqi}). We find that spectral diffusion is the main limitation to the visibility while the effect of pure dephasing is negligible. As spectral diffusion is slow, it can be mitigated dynamically by more frequent or stricter resonance checking. Additionally, Sn ion implantation at lower dose and energy or tailored implantation methods \cite{bushmakinTwoPhotonInterference2024} could yield devices with less charge noise.
For the remote entanglement experiments below, we set $\left|\Delta t\right|!{max}$ to \qty{10}{\ns} (dotted line in Fig.~\ref{fig:tpqi}(d)), yielding a visibility of $\approx \num{0.86}$ for a \qty{16}{\%} reduction in two-photon detection probability and hence data rate.

\section{Heralded Remote Entanglement}
We generate remote entanglement between the two spin qubits using the Barret-Kok \cite{barrettEfficientHighfidelity2005} protocol. This protocol starts by generating entanglement between the spin qubit and a photonic time-bin qubit at each node in the following way. First, the spin qubit is prepared in a balanced superposition state, followed by an optical $\pi$-pulse that selectively excites the $\ket{0}$-state to create a photon in the "early" photonic mode ($\ket{E}$). After a $\pi$-rotation of the spin, a second round of state-selective excitation yields a photon in the "late" photonic mode ($\ket{L}$), resulting in the joint state $\ket{\psi} = (\ket{0}\ket{L}+ \ket{1}\ket{E})/\sqrt{2}$.

We benchmark this first step of the protocol by repeatedly generating the spin-photon state and reading out the spin state once the photon is detected in the first or second time bin. Fig.~\ref{fig:entanglement}(a) displays the resulting spin measurement outcomes as a function of the photon detection time bin. We find near-perfect correlations for both nodes, evidencing the desired control over the spin-photon interface.

In the next step of the protocol, the photons from the two nodes are interfered on a central beamsplitter. Finally, successful spin-spin entanglement is heralded by detection of one photon in each of the two time bins. 

We implement this protocol using the experimental sequence shown in Fig.~\ref{fig:entanglement}(b).
Once a heralding signal is received, the qubits are measured in one of the cardinal bases to obtain the two-qubit correlators $\langle XX\rangle$, $\langle YY\rangle$ and $\langle ZZ\rangle$. Depending on the photon detection pattern, either $\ket{\Psi^+}$ (same detector clicks twice) or $\ket{\Psi^-}$ (each detector clicks once) is generated. In order to deliver the state $\ket{\Psi^-}$ independent of the heralding pattern, a conditional Pauli correction is performed in real time via a compiled $Z$-gate on Bob.

The joint two-qubit measurement outcomes split up per basis are shown in Fig.~\ref{fig:entanglement}(c). All three measurement bases exhibit strong anti-correlations as expected for $\ket{\Psi^-}$.%The probabilities are corrected for readout errors (*ref to appendix*). 
We calculate the resulting correlator values which yield a state fidelity of $\mathcal{F} = (1- \langle XX\rangle - \langle YY\rangle- \langle ZZ\rangle)/4=  \qty{0.730(11)}{}$, confirming the successful generation of remote spin-spin entanglement.

The grey bars in Fig.~\ref{fig:entanglement}(c) indicate the expected values based on an analytical model capturing the major infidelity sources of our experiment (Appendix \ref{app:model_entanglement}). We find good quantitative agreement between the model and the data. 
The major infidelity contributions are the photon interference visibility, imperfect spin selectivity of the optical $\pi$-pulses and double excitation.
The latter two errors can be reduced significantly by increasing the magnetic field (Appendix \ref{app: magnetic field}), leading to a larger separation of the optical transitions and hence increased selectivity. Additionally, improved optical pulse shaping can further suppress these errors. 

The entanglement generation in this two-photon protocol has a success probability per attempt of $(p_A \cdot p_B)/2$. Here, $p_A$ ($p_B$) is the probability that a resonant photon is emitted by Alice (Bob) and detected at the heralding station, which in this experiment is $\qty{0.23(1)}{\%}$ ($\qty{0.30(2)}{\%}$), mainly limited by photon collection efficiency. 
This can be significantly improved in the near term by in-contact coupling of the tapered optical fiber to the nanophotonic waveguide \cite{zengCryogenicPackaging2023}, by fabricating devices from thin-film diamond with reduced scattering due to bottom-surface roughness \cite{guoDirectbondedDiamond2024} and by incorporating a reflector at the waveguide end \cite{parkerDiamondNanophotonic2024}. With these improvements, resonant photon collection efficiencies of tens of percent are feasible. Additionally using the loss-resilient single-photon protocol at an entanglement fidelity cost of 0.05 is projected to yield entanglement success probabilities at percent level (Appendix~\ref{app: ideal entanglement prob estimate}). At system level, the success probability can then be further increased by orders of magnitude by time and frequency multiplexing of the entanglement generation.

\section{Outlook}
We have demonstrated a quantum network link based on diamond tin-vacancy centers integrated into broadband nanophotonic waveguides. By removing the need for resonant optical structures and the associated complex fabrication and tuning, these results open up a near-term pathway for scaling network nodes. In particular, photonic integration of these waveguide devices \cite{wanLargescaleIntegration2020,liHeterogeneousIntegration2024} combined with in-situ frequency tuning \cite{clarkNanoelectromechanicalControl2024, brevoordLargerangeTuning2025} could enable thousands of qubits to be connected on-chip with high yield. By operating these in sync for temporal and frequency multiplexing, while applying quantum frequency conversion for low-loss photon transmission, kilohertz entangling rates at metropolitan scale come within reach (Appendix~\ref{app: ideal entanglement prob estimate}). By additionally leveraging qubit storage enabled by the long-lived spin states \cite{beukersControlSolidState2025, harrisHighFidelityControl2025, reschHighFidelityControl2026}, buffers with significant numbers of entangled pairs can be maintained, effectively turning the probabilistic entanglement generation into a deterministically available network resource. Such entanglement buffers enable new applications including low-latency coordination \cite{arunFasterthanlightCoordination2025,dasilvaEntanglementImproves2025} and device-independent randomness generation \cite{kavuriTraceableRandom2025}, and establish a general-purpose infrastructure for networking remote quantum devices.

\section{Acknowledgements}
The authors would like to thank Silvia Ruffieux and Patrick Maletinsky from Department of Physics in University of Basel for providing access to and support with their annealing oven, Mariagrazia Iuliano, Niv Bharos and Conor Bradley for fruitful discussions, and Pepijn Habing and Tim Turan for early device characterization. 

We acknowledge financial support from the joint research program “Modular quantum computers” by Fujitsu Limited and Delft University of Technology co-funded by the Netherlands Enterprise Agency under project number PPS2007, from the Dutch Research Council (NWO) through the Spinoza prize 2019 (project number SPI 63-264), from the Dutch Ministry of Economic Affairs and Climate Policy (EZK) as part of the Quantum Delta NL programme, and from the Quantum Internet Alliance through the Horizon Europe program (grant agreement No. 101080128).

\section{Author contributions}
C.W., H.K.C.B, and R.H. devised the experiment. C.W., T.D., H.K.C.B., and A.M.S. prepared the experimental apparatus. N.C. fabricated the device. N.M. developed software for data acquisition. C.W., T.D., and A.M.S. carried out the experiments and analyzed the data. C.W., T.D., A.M.S., and R.H. wrote the main manuscript with input from all authors. C.W., T.D., H.K.C.B., A.M.S., N.C., and R.H. wrote the supplementary materials. R.H. supervised the research.

\section{Data, code and materials availability}
The data that support the findings in this article are openly available \cite{4TUrepository}. No physical materials were generated in this work.

\newpage
\appendix

\section{Device}
\begin{figure*}
	\centering
	\includegraphics[width=\textwidth]{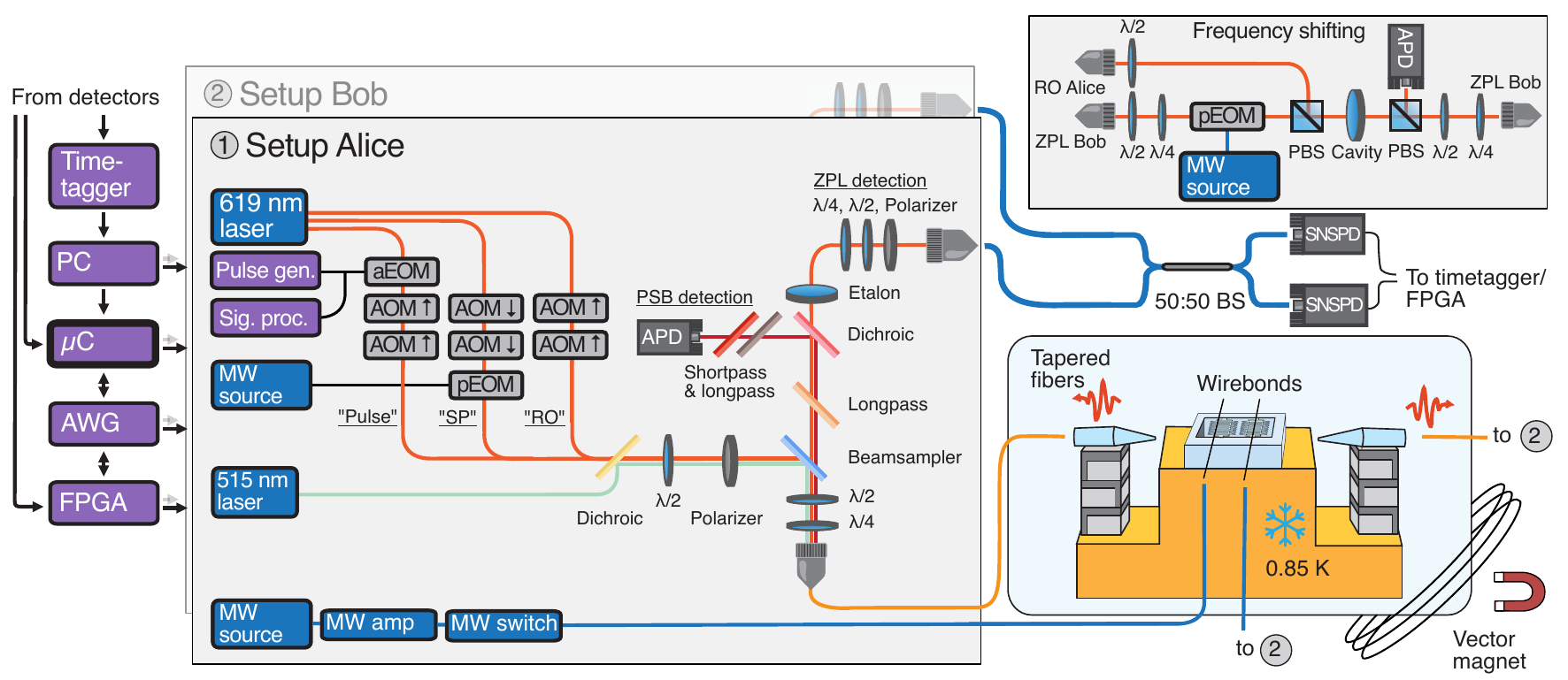}
	\caption{\textbf{Experimental setup:} The components of the Alice setup are shown. The Bob setup is a replica of the Alice setup and only indicated. "RO": readout, "SP": spinpump and "Pulse" refer to the respective resonant laser paths. A detailed description of the components can be found in Appendix~\ref{app:setup}.}
	\label{fig:setup}
\end{figure*}

The devices are realized through two primary fabrication steps: (1) fabrication of SnV centers and (2) fabrication of nanophotonic waveguides.

The fabrication of SnV centers in diamond is based on the approach of ion implantation and high-temperature low-pressure annealing \cite{codreanuAboveUnityCoherent2025, codreanuDiamondNanophotonic2025}, whereas the fabrication of nanophotonic waveguides is based on the crystal-dependent quasi-isotropic undercut etch fabrication process \cite{khanalilooHighQMonolithic2015,mouradianRectangularPhotonic2017,rugarNarrowLinewidthTinVacancyCenters2020,codreanuAboveUnityCoherent2025}.

Fabrication of SnV centers starts with an Element Six $\qty{4}{\mm}\times\qty{4}{\mm}\times\qty{500}{\um}$ $\{100\}$ surface-oriented electronic grade diamond plate. First, the sample undergoes a thorough inorganic wet clean in Hydrofluoric Acid (HF, \qty{40}{\%} concentration) for \qty{20}{\min} at room temperature, followed by a Piranha mixture (ratio 3:1 of \ce{H2SO4 (95\%)}~:~\ce{H2O2 (31\%)}) for \qty{20}{\min} at \qty{80}{\degreeCelsius}. Next, we proceed with an Inductively Coupled Plasma - Reactive Ion Etching (ICP-RIE) strain-relief etch in \ce{Ar/Cl_2} plasma to remove the residual polishing-induced strain from the substrate surface, thereby etching a diamond layer of $\approx\qty{5}{\um}$. This is then followed by an ICP-RIE etch in \ce{O_2} plasma ( $\approx\qty{6}{\um}$ etched diamond layer), to remove residual chlorine contamination from the previous etching step \cite{rufCavityenhancedQuantum2021}.

The sample is then inorganically cleaned in a Piranha mixture ($\simeq$ \qty{20}{\min} at \qty{80}{\degreeCelsius}) and implanted with \ce{^{120}Sn^{++}} ions, with the following implantation parameters: \ang{7}~angle of implantation with respect to normal surface, ion fluence dose \qty{1e11}{\text{ions}/cm^2}. The implantation energy is \qty{350}{\keV}, which is predicted to yield an implantation depth of \qty{88(14)}{\nm} from Stopping Range of Ions in Matter (SRIM) simulations.

Before the SnV activation via high temperature low pressure annealing (\qty{1100}{\degreeCelsius} for 10 hours \cite{rufOpticallyCoherent2019, codreanuDiamondNanophotonic2025}), the sample undergoes a wet tri-acid mixture inorganic clean  (ratio 1:1:1 of \ce{HNO3}$(\qty{65}{\%})$ : \ce{HClO4}$(\qty{70}{\%})$ : \ce{H2SO4}$(\qty{>99}{\%})$) in a reflux configuration setup at \qty{120}{\degreeCelsius} for \qty{70}{\min}, to remove any organic contamination. After the high temperature annealing, the sample undergoes a sequential double tri-acid mixture inorganic clean at \qty{120}{\degreeCelsius} for \qty{120}{\min} each. This is then followed by a short ICP-RIE \ce{O_2} etch, ensuring a full removal of the non-diamond graphitic thin film layer that formed during the annealing step. The additional short plasma etch brings the SnV centers slightly closer to the surface than designed. 

Due to insufficient activation of SnV centers during the first annealing, a second high temperature low pressure annealing round is performed (\qty{1100}{\degreeCelsius} for a duration of 4 hours), employing our surface-protected annealing method \cite{codreanuAboveUnityCoherent2025, codreanuDiamondNanophotonic2025}. The second annealing round is preceded by a sequential double tri-acid mixture inorganic clean at \qty{120}{\degreeCelsius} for \qty{120}{\min} each. After annealing, we proceed with a tri-acid mixture inorganic clean (for a duration of \qty{180}{\min} at \qty{120}{\degreeCelsius}) to remove any non-diamond graphitic thin film layer (on non-covered edges of the diamond substrate), as well as to oxygen terminate the overall diamond surface. Before proceeding with the fabrication of nanophotonic waveguides, the sample is characterized at \qty{5}{\K} in a closed-cycle cryostat to confirm the successful activation of SnV centers.
 
The fabrication of nanophotonic devices in this work follows similar process parameters detailed in Ref.~\cite{codreanuAboveUnityCoherent2025}. However, in this work, the total duration of the crystal-dependent quasi-isotropic undercut etch differs from our previous work:
here the full release and upward quasi-isotropic etch is executed in 3 consecutive separate etch steps for a total duration of \qty{67}{\min}. 

The geometry of the fabricated devices was inspected with a scanning electron microscope on a representative subset of array of devices and confirmed to be in compliance with the design requirements. The extensive structural integrity of all devices on the sample is then assessed by optical microscopy (Keyence digital microscope system).
We categorize the devices in 3 classes: (1) structurally conforming (8,250), (2) showing fabrication defects (68) and (3) physically broken by chip handling (18). Of the 8,336 patterned waveguide devices (by chip design), for the calculation of the structural yield, we consider the 8,250 structurally conforming devices, resulting in a structural yield of $\qty{99}{\%}$.

\section{Setup} \label{app:setup} \label{sec:setup}
The experimental setup is shown schematically in Fig.~\ref{fig:setup}. A cryostat (Bluefors LD250He, no dilution unit, circulation to \qty{1}{\K} pot with Helium-4) cools a diamond with nanophotonic waveguides and SnV centers to a base temperature of \qty{850}{mK}. Magnetic fields are created with a three-axis vector magnet (AMI). Two concatenated \textit{4f}-lens systems (not shown) allow optical imaging. Two tapered optical fibers (Thorlabs S600-HP, taper angle $\approx$$\ang{3}$) on nanopositioning stages (Attocube, ANPx101 and ANPz102) function as interfaces to the nanophotonic waveguides in lensed-fiber configuration. Microwaves are delivered via wirebonds. 

Two separate lasers are used for Alice and Bob, respectively. A \qty{515}{\nm} laser (Hübner Photonics, Cobolt 06-03-MLD515) is used for charge initialization and a \qty{619}{\nm} laser (Toptica Photonics AG, DL Pro SHG) for resonant excitation. The laser frequency of the resonant laser is stabilized to a wavemeter (HighFinesse, WS-8) and the laser light is split into three paths. The 'readout' path passes two up-shifting acousto-optic modulators (AOM, Gooch \& Housego, Fiber-Q 637nm) and is used for resonant readout. The 'spinpump' laser path passes two down-shifting acousto-optic modulators (AOM, Gooch \& Housego, Fiber-Q 637nm) as well as a phase electro-optic modulator (pEOM, Jenoptik, PM635). The pEOM is driven by a microwave (MW) source (Rohde \& Schwarz, SMBV100A) and the MW frequency is chosen such that the 1st order EOM sidebands are resonant with 
one of the spin-conserving transitions for spinpumping. The down-shifting AOMs ensure that the zero order sideband of the pEOM is not at the same frequency as the readout and pulse paths.
The third path, 'pulse', is used for creating short optical pulses and consists of one electro-optic modulator (aEOM, Jenoptik, AM635) and one up-shifting AOM (Crystal Technologies Inc.) in free-space double-pass configuration. The bias of the aEOM is stabilized with a signal processor (RedPitaya, STEMlab 125-14) and ns-short electrical pulses are applied via a home-built pulse generator. 

For the results in Fig.~\ref{fig:tpqi} and Fig.~\ref{fig:entanglement}, both pulse paths are connected to the same resonant laser. This is done to minimize the frequency difference between the pulsed excitations on the two emitters in the entanglement protocol without the need for a frequency lock between the lasers. This is crucial for the experiment since both lasers are frequency locked to the signal of a wavemeter, leading to shot-to-shot fluctuations of the frequency difference between the lasers on the order of \qty{1}{\MHz} (larger than the inverse time-bin duration), which would randomize the phase of the entangled state from shot to shot.

\begin{figure*}
    \centering
    \includegraphics{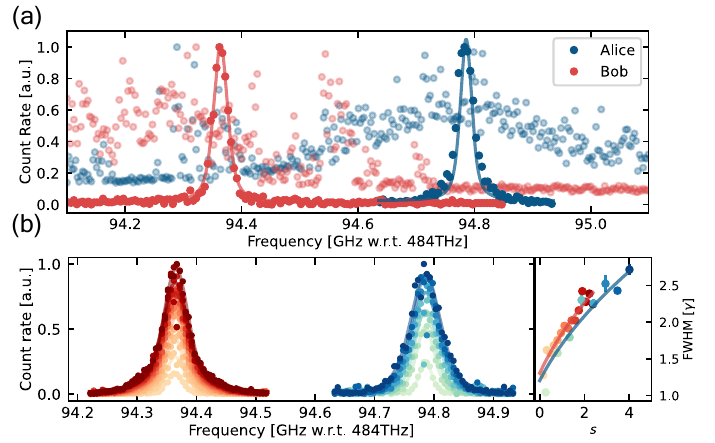}
    \caption{\textbf{Characterization of the optical transitions.} \textbf{(a)} Measurement of the photoluminescence upon resonant excitation while varying the frequency of the excitation. The transparent data represent the measurement where the emitters are prepared by a single repump pulse before each resonant excitation. The fitted solid data is the same experiment, but when each resonant excitation is preceded by a CR-check. \textbf{(b)} Left: Same measurement as the solid data in (a), but now for varying excitation powers. Right: Fitted Full-Width at Half Maximum (FWHM) of the measured lines for the different excitation powers (expressed in $s=P/P_\text{s}$ with $P_\text{s}$ the saturation power). Data is fitted with a power broadening model.}
    \label{fig: supp - PLEs}
\end{figure*}

All laser paths are combined via dichroic mirrors (Thorlabs, DMLP567R) and beamsplitters and routed via the optical fiber into the cryostat and to the SnVs. A set of quarter- and half-waveplates (Thorlabs, WPQ05M-633/WPH05M-633) in motorized rotation mounts (Thorlabs, K10CR1) set the polarization of the laser light.

The light emitted by the SnVs is collected via the same optical fiber and routed outside of the cryostat. A beamsampler (Thorlabs, BSF10-A) separates excitation and collection path. A long-pass filter (Thorlabs, FELH0600) filters reflections of the \qty{515}{\nm}  laser light to avoid blinding of the single photon detectors. Subsequently, the emitted light is separated into PSB and ZPL via a dichroic mirror (Semrock, FF625-SDi01). 

The PSB photons are spectrally filtered further via a tunable long-pass filter (Semrock, TLP-01-628) to remove reflections of the resonant laser light and a short-pass filter at \qty{700}{\nm} (Thorlabs, FESH0700) to filter auto-fluorescence of the optical fiber. The photons are coupled into a multi-mode fiber (Thorlabs, FG050LGA) and routed to an avalanche photon detector (LaserComponents, COUNT-R). 

The ZPL photons are filtered by an etalon (LightMachinery Inc, custom device) with a full-width at half maximum of around \qty{40}{\GHz}. Another set of waveplates and a polarizer with the same components are used to remove reflected laser light. 
The ZPL photons are then coupled into a polarization maintaining fiber connected to a balanced beamsplitter (Evanescent Optics, custom device). The two output ports are routed to superconducting nanowire single photon detectors (PhotonSpot, custom device).

For the frequency shifting of the ZPL photons used in Sec. \ref{app:AddTPQI}, we additionally use frequency shifting via a free-space phase EOM. The ZPL photons from Bob are routed via a polarization-maintaining fiber to a separate breadboard. After passing a set of quarter- and half-waveplates (Thorlabs, WPQ05M-633/WPH05M-633), the photons are frequency-shifted by a free-space phase EOM (Qubig, TWP10M1-VIS) to the 1st-order sidebands. The microwave signal is supplied by a microwave source (Rohde \& Schwarz, SMBV100A). The ZPL path is combined by a polarizing beamsplitter (Thorlabs, PBS251) with a tap-off from the readout path of the Alice laser, which is used to lock a cavity at the frequency of the emitted photons of Alice. The cavity (LightMachinery Inc) has a full-width at half maximum of around \qty{130}{\MHz} and is locked to maximum transmission by a micro-controller (see below). We find that the passive stability of the cavity is good enough to relock it only once the transmission drops under a threshold. Focussing lenses (not shown) are used both for both the phase EOM and the cavity and to enable optimal mode-matching and transmission. The locking light is split by another PBS and detected by an avalanche photon detector (LaserComponents, COUNT-R). The ZPL light passes another set of waveplates and is then coupled into the central 50:50 beamsplitter. The overall transmission through the breadboard (including fiber coupling) is $\sim$$\qty{46}{\%}$ whereas $\sim$$\qty{30}{\%}$ of the incoming light is shifted to one of the 1st order sidebands, yielding an overall efficiency for frequency-shifting of $\sim$$\qty{14}{\%}$. 

Microwave fields for spin control are generated by a source (Rohde \& Schwarz, SGS100A), amplified (Amplifier Research, 40S1G4 / MiniCircuits, ZHL-50W--52-S+) and routed to the cryostat. A homebuilt microwave switch disconnects the transmission into the cryostat whenever no signal needs to be supplied and a set of high-pass filters and DC-blocks (MiniCircuits) suppresses low-frequency noise. Inside the cryostat, the signal is routed via superconducting coaxial cables (CoaxCo/QuantumCoax) to a printed circuit board enclosing the diamond. A suspended bondwire, distanced approximately \qty{100}{\um} above the devices, delivers the signal to the SnVs. 

The electronic output signal of the photon detectors is recorded by a timetagger (PicoQuant, HydraHarp400) and later retrieved by a measurement computer. The signal is additionally routed to a counter module in the microcontroller (description below), enabling real-time processing of photon events required for charge-resonance checks, as well as an FPGA (T-Rex Junior, JigSaw B.V.). The FPGA allows real-time operation for conditional MW pulse switching, heralding successful entanglement generation, and Pauli correction depending on the detection pattern. 

All control signals required to shape the laser and microwave pulses are supplied by a combination of a micro-controller (Jäger Messtechnik, ADwin Pro II) and arbitrary waveform generator (Zurich Instruments, HDAWG).

The experiments are prepared in a homebuilt software framework which supplies the required logical sequences to the ADwin and HDAWG and retrieves measured data from the ADwin counters as well as the timetagger. All interfaces to hardware components are established via the quantum measurement infrastructure QMI \cite{raaQMIQuantumMeasurement2023}. 

\section{Magnetic Field Tuning} \label{app: magnetic field}
As mentioned in the main text, the magnetic field in this work is used primarily to overlap the optical transitions of two SnV centers with different optical transition frequencies at zero field. By applying a magnetic field, the optical frequencies of the two spin-conserving transitions become non-degenerate. The frequency difference between the optical frequencies is determined by the magnetic field amplitude and direction. From the zero-field difference between the optical frequencies of the two SnV centers, we can directly calculate a set of magnetic field settings (amplitude and direction) that overlaps a combination of their optical transitions.

The magnetic vector field is defined in the coordinates of the lab frame. We define the z-axis to be orthogonal to the sample, pointing upwards. We then define the x-axis to be in the sample plane, orthogonal to the directions of the waveguides, which means that the y-axis is defined along the waveguides. From this we define a spherical coordinate system through the coordinates $(r,\theta,\phi)$.

From the diamond crystal structure, we know that four orientations of the SnV dipole moment are possible on our sample. Expressed as combinations of $(\theta, \phi)$, we define these as: $\text{x} \equiv (\ang{54.7}, \ang{0})$, $\text{y}\equiv(\ang{54.7}, \ang{90})$, $\text{-x} \equiv (\ang{54.7}, \ang{180})$ and $\text{-y} \equiv (\ang{54.7}, \ang{270})$. The Alice emitter has the "-y" orientation, while the Bob emitter has the "-x" orientation. Given the fact that we use two emitters with different orientations in a single globally defined magnetic field, we must operate both emitters at misaligned magnetic fields.

From the homogeneous linewidths of the optical transitions of the emitters, we get a range of possible magnetic fields that would sufficiently overlap the optical emission frequencies of the two emitters. This set of possible field settings is further restricted by the maximum spin splitting for which the microwave transmission is sufficient for spin control. Within these restrictions, the final field setting is chosen to balance the product of optical cyclicity and PSB collection efficiency for both SnV centers. This is done to operate both in a regime with similar single shot readout probabilities. The magnetic field that is used for the experiments in the main text is $|\Vec{B}|=\qty{117}{mT}$, $\theta=\ang{46}$ and $\phi=\ang{250}$.
\begin{figure*}
    \centering
    \includegraphics{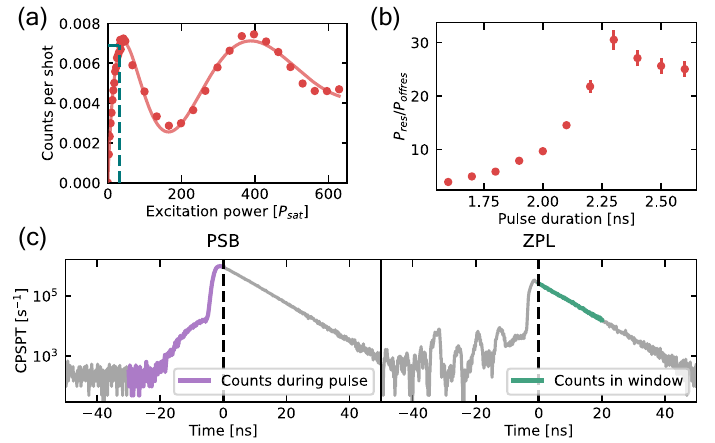}
    \caption{\textbf{Optical pulse optimisation for the Bob emitter.} \textbf{(a)} Measurements of the counts per shot (CPS) in the PSB for varying powers of the optical pulse. Data is fitted to an optical Rabi model. \textbf{(b)} Measured probability of off-resonant excitation for varying optical pulse durations. Pulse powers are adjusted for each of the data points to achieve the same population inversion for all pulse durations. \textbf{(c)} Estimation of the double excitation probability. Left: Trace of the measured photons in an optical pulse experiment in the PSB. The purple part of the trace defines the time window during which we count photons for double excitation. Right: Photon counts from the same measurement, but measured in the ZPL. The green part of the trace indicates the measurement window after the pulse.}
    \label{fig: supp - optical pulses}
\end{figure*}

\section{Additional Characterization Data}
\subsection{Characterization of the Frequency Stability}
\label{app:ple}
As outlined in Appendix~\ref{app: statistics}, our approach to finding SnV centers in a waveguide starts with the appearance of fluorescence in a photoluminescence excitation (PLE) measurement. In such a measurement, resonant excitation is always preceded by off-resonant (\qty{515}{\nm}) excitation to prevent ionization of the SnV center. The observed count rates from such a measurement for both SnV centers is shown by the transparent data in Fig.~\ref{fig: supp - PLEs}(a). It is very clear that the resulting fluorescence from such a measurement has a very broad frequency range as a result of pseudo-random charge reshuffling from the off-resonant excitation, combined with slow spectral diffusion contributions.

For the generation of indistinguishable photons it is crucial that the emission frequencies of both tin vacancy centers are (1) stable and (2) known apriori. In this work, we employ the technique of Charge-Resonance (CR) checks \cite{brevoordHeraldedInitialization2024} in our sequences in an effort to meet the frequency stability requirements for our experiments. To demonstrate the effectiveness of this technique, we also measure the PLE spectrum of our emitters when replacing the off-resonant excitation by a CR check. The figure shows how these CR checks are able to significantly reduce the linewidths of the optical transitions for both SnV centers. Next to this, the post-CR-check frequency of the optical transition of the SnV is automatically aligned to the frequency of the CR-check probe laser.

We aim to get a precise estimate of the linewidth after passing the CR check, which is the relevant linewidth for the quantum network operation. Fig.~\ref{fig: supp - PLEs}(b) shows the resonant excitation spectra of both emitters as a function of excitation power. All resonant excitations here are preceded by a CR check. By fitting the relation between the fitted Full-Width at Half Maximum (FHWM) and the excitation power, we find the linewidth of both emitters in the limit of zero power broadening: $\Gamma_0^{\text{A}} = \qty{27(4)}{\MHz} = 1.3(2)\,\gamma!A$ and $\Gamma_0^{\text{B}} = \qty{29.7(1.4)}{\MHz} = 1.44(7)\,\gamma!B$, where $\gamma!{A/B}=\frac{1}{2\pi\tau!{A/B}}$ is the lifetime-limited linewidth. 

\subsection{Characterization of the Optical Pulses}
\label{app:pulses}
\label{app:Ramsey}

As outlined in Appendix~\ref{app: magnetic field}, the magnetic field amplitude and direction are fully determined by (1) the frequency difference between the emitters at zero field, (2) the (different) orientations of the symmetry axes of the two SnV centers and (3) the maximum spin splitting for which we get sufficient microwave transmission through the setup. Effectively, this means that the magnetic field regime is not optimized for spin-selective resonant pulsed excitation. This section discusses the techniques we employ to optimize the optical pulse delivery for the given magnetic field regime in which we operate the emitters.

Firstly, Fig.~\ref{fig: supp - optical pulses}(a) shows a measurement of the counts per shot (CPS) after a \qty{2.3}{\ns} resonant optical pulse measured in the PSB path and integrated over the same \qty{20}{\ns} window as defined in Fig.~\ref{fig:characterization}(a). We perform this measurement for a range of excitation powers in the pulse, which we control by changing the optical transmission through a double-pass AOM. From the fitted Rabi oscillations, we obtain the optical power that is required to do a full inversion of the population from ground- to excited state. To minimize the double excitation probability, while maintaining a high excitation probability, we calibrate the optical pulses to make a $0.85\pi$ rotation of the population, which we refer to as an optical $\pi$-pulse in this work for simplicity.

Next, we aim to minimize the spectral amplitudes of the pulse which overlap with the spin-conserving optical transition of the spin state we are not addressing. In our experiment, the SnV center labeled Bob has the lowest optical splitting of $\Delta f = \qty{372}{\MHz}$. If we assume the temporal shape of the excitation pulses to be perfectly square, then a pulse of duration $T$ will have a spectral intensity profile $S(\omega)$ that follows
\begin{equation} \label{eq: Opt pulse spectral shape}
    S(\omega) \sim \text{sinc}^2 \left( \frac{(\omega - \nu)T}{2} \right),
\end{equation}
where $\nu$ is the central frequency of the optical mode that constitutes the pulse. The spectral distribution of a square pulse thus has zero-crossings at distances of $\Delta \omega=\frac{2\pi}{T}$ from the central frequency. This means that we can eliminate the off-resonant excitation probability by choosing a pulse duration of $T=\frac{1}{\Delta f}$. In practice, perfectly square optical pulses are not feasible due to finite fall/rise times of both optical and electronic components. To obtain a measure of the off-resonant excitation probability, we perform an experiment with interleaved measurements of the single photons in the PSB after optical pulses when preparing the spin in both the resonant and off-resonant states. The ratio of the measured photons in both experiments is plotted in Fig.~\ref{fig: supp - optical pulses}(b) as a function of the duration of the optical pulse. The figure clearly shows how the broadness of the frequency spectrum of shorter optical pulses causes significant off-resonant excitation probability, making them unusable for protocols that require spin-selective excitation. From the measurement in Fig.~\ref{fig: supp - optical pulses}(b), we decide on an optical pulse duration on the Bob emitter of $T = \qty{2.3}{\ns} = 0.32\tau!B$. 

The minimum duration of the optical pulse is thus mostly fixed from the optical splitting of the spin-conserving transitions. However, the fidelity of the heralded entanglement is also affected by instances of double excitation (see Appendix~\ref{app:model_entanglement}). Double excitation describes instances where a spontaneous emission event occurs during the optical pulse, after which the emitter is re-excited and undergoes another cycle of spontaneous decay. We measure the double excitation probability by finding the number of occurrences where we measure a photon during the optical pulse, as well as in the \qty{20}{\ns} time window after the optical pulse. As it is impossible to separate SnV emitted photons from laser reflections in the ZPL during the pulse, we measure the emitted photons during the pulse only in the PSB and the photons in the measurement window only in the ZPL. The data of this measurement is shown in Fig.~\ref{fig: supp - optical pulses}(c), where the colored sections of the data represent the time windows in which photons are counted. From this, we can subsequently estimate the double excitation probability with
\begin{equation} \label{eq: Double Excitation}
    p_{de} = \frac{N_\text{coin}}{(1 - \eta_\text{DW})\eta_\text{PSB}N_\text{ZPL}}.
\end{equation}
Here $N_\text{coin}$ is the number of repetitions where both a PSB photon and a ZPL photon in the above described windows is measured. $N_\text{ZPL}$ is the total number of measured photons in the time window in the ZPL, $\eta_\text{PSB}$ is the collection efficiency in the PSB collection path and $\eta_\text{DW}=0.57$ is the Debye-Waller factor of photons emitted in the ZPL and PSB. 

\subsection{Ramsey Measurements}
\begin{figure*}
    \centering
    \includegraphics{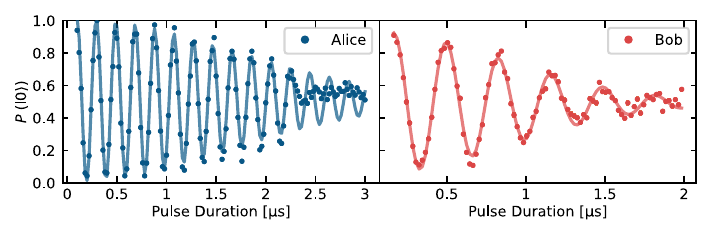}
    \caption{\textbf{Ramsey measurements probing the free induction decay of the spin qubit coherence}. The coherence oscillates with the detuning $\Delta$ of the MW drive from the electron transition and shows Gaussian decay.}
    \label{fig: supp - Ramsey}
\end{figure*}
\textbf{\label{app:ODMR}
\begin{figure*}
    \centering
    \includegraphics[width=0.8\columnwidth]{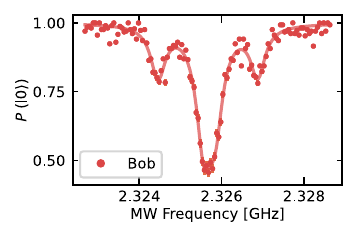}
    \caption{\textbf{ODMR measurements of the Bob node.} The triple peak structure stems from a nearby strongly coupled $^{13}$C and the two electron-spin flipping transitions appearing nearly degenerate.}
    \label{fig: supp - ODMR}
\end{figure*}}
In addition to the spin coherence measurements in Fig.~\ref{fig:characterization}(c), Fig.~\ref{fig: supp - Ramsey} shows Ramsey measurements with a detuned drive. The spin coherence shows a Gaussian decay $ \propto \exp{(-(t/T_2^*)^2)}$ by a slowly varying nuclear spin bath. We find an inhomogeneous dephasing time $T_2^*=$~\qty{2.32(7)}{\us} for the Alice node and $T_2^*=$~\qty{1.22(3)}{\us} for the Bob node. The applied magnetic fields are $|\vec B|$= \qty{163}{mT}, $\theta$ = \ang{45}, $\phi$ = \ang{225}  for Ramsey measurements on Alice and $|\vec B|$= \qty{117}{mT}, $\theta$ = \ang{46}, $\phi$ = \ang{250} for Ramsey measurements on Bob.

We find that the optically detected magnetic resonance (ODMR) measurements on Bob show an additional structure (Fig.~\ref{fig: supp - ODMR}), in line with a single strongly coupled $^{13}$C nuclear spin. The two overlapping inner transitions are the electron-spin flipping (single quantum) transitions and the outer peaks are the zero- and double-quantum transitions. We center the microwave frequency in all experiments between the two center transitions and ignore the $^{13}$C nuclear spin. By sweeping the magnetic field magnitude along the direction $\theta$ = \ang{46}, $\phi$ = \ang{250} (not shown), we find that the splitting matches the gyromagnetic ratio of $^{13}$C and estimate the hyperfine coupling to be $|\mathbf A/2 \pi|=$ \qty{0.25}{\MHz}.

\section{Additional Details on Experimental Procedure}
\label{app:procedure}
\renewcommand{\arraystretch}{1.3}
\begin{table*}
    \centering
    \caption{\textbf{Power and duration settings for all optical control pulses.} Overview of the power and duration settings used for the optical control pulses in the figures in this work. Resonant optical powers are expressed in terms of saturation power $P!s$, the value of which is calibrated through independent measurements Uncertainties in readout duration reflect small fluctuations in the calibrated readout length throughout the experiment. $^{1}$Powers for the optical pulse are estimations based on the EOM on/off ratio.}
    \label{tab: optical settings}
    \begin{tabular}{|c|c|c|c|c|c|c|}
        \hline
         &  & CR Repump & CR Probe & Reset & Readout & Optical pulse$^1$ \\ \hline
        \multirow{2}{*}{Fig.~\ref{fig:characterization}(a)} & Alice & $\qty{10}{\uW}$, $\qty{50}{\us}$ & $\num{0.14}P!s$, $\qty{100}{\us}$ & $\num{0.97}P!s$, $\qty{100}{\us}$ & \multirow{2}{*}{N/A} & $\num{315}P!s$, $\qty{1.59}{\ns}$ \\ \cline{2-5} \cline{7-7} 
         & Bob & $\qty{1}{\uW}$, $\qty{10}{\us}$ & $\num{0.19}P!s$, $\qty{80}{\us}$ & $\num{0.40}P!s$, $\qty{80}{\us}$ &  & $\num{55}P!s$, $\qty{2.3}{\ns}$ \\ \hline
        \multirow{2}{*}{Fig.~\ref{fig:characterization}(b)} & Alice & $\qty{0.5}{\uW}$, $\qty{10}{\us}$ & $\num{0.18}P!s$, $\qty{100}{\us}$ & $\num{1.31}P!s$, $\qty{100}{\us}$ & $\num{1.05}P!s$, $\qty{15}{\us}$ & \multirow{2}{*}{N/A} \\ \cline{2-6}
         & Bob & $\qty{20}{\uW}$, $\qty{10}{\us}$ & $\num{0.13}P!s$, $\qty{80}{\us}$ & $\num{0.54}P!s$, $\qty{80}{\us}$ & $\num{0.72}P!s$, $\qty{14}{\us}$ &  \\ \hline
        \multirow{2}{*}{Fig.~\ref{fig:characterization}(c)} & Alice & $\qty{0.5}{\uW}$, $\qty{10}{\us}$ & $\num{0.18}P!s$, $\qty{100}{\us}$ & $\num{1.31}P!s$, $\qty{100}{\us}$ & $\num{1.05}P!s$, $\qty{17}{\us}$ & \multirow{2}{*}{N/A} \\ \cline{2-6}
         & Bob & $\qty{20}{\uW}$, $\qty{10}{\us}$ & $\num{0.13}P!s$, $\qty{80}{\us}$ & $\num{0.54}P!s$, $\qty{80}{\us}$ & $\num{0.72}P!s$, $\qty{14}{\us}$ &  \\ \hline
        \multirow{2}{*}{Fig.~\ref{fig:tpqi}} & Alice & $\qty{0.5}{\uW}$, $\qty{10}{\us}$ & $\num{0.15}P!s$, $\qty{100}{\us}$ & $\num{1.04}P!s$, $\qty{100}{\us}$ & \multirow{2}{*}{N/A} & $\num{355}P!s$, $\qty{1.59}{\ns}$ \\ \cline{2-5} \cline{7-7} 
         & Bob & $\qty{1}{\uW}$, $\qty{10}{\us}$ & $\num{0.21}P!s$, $\qty{80}{\us}$ & $\num{0.45}P!s$, $\qty{80}{\us}$ &  & $\num{50}P!s$, $\qty{2.3}{\ns}$ \\ \hline
        \multirow{2}{*}{Fig.~\ref{fig:entanglement}(a)} & Alice & $\qty{0.5}{\uW}$, $\qty{10}{\us}$ & $\num{0.17}P!s$, $\qty{100}{\us}$ & $\num{1.20}P!s$, $\qty{100}{\us}$ & $\num{0.96}P!s$, $\qty{15}{\us}$ & $\num{265}P!s$, $\qty{1.59}{\ns}$ \\ \cline{2-7} 
         & Bob & $\qty{1}{\uW}$, $\qty{10}{\us}$ & $\num{0.17}P!s$, $\qty{80}{\us}$ & $\num{0.74}P!s$, $\qty{80}{\us}$ & $\num{0.99}P!s$, $\qty{12}{\us}$ & $\num{70}P!s$, $\qty{2.3}{\ns}$ \\ \hline
        \multirow{2}{*}{Fig.~\ref{fig:entanglement}(c)} & Alice & $\qty{0.5}{\uW}$, $\qty{10}{\us}$ & $\num{0.17}P!s$, $\qty{100}{\us}$ & $\num{1.23}P!s$, $\qty{100}{\us}$ & $\num{0.99}P!s$, $\qty{16(2)}{\us}$ & $\num{240}P!s$, $\qty{1.59}{\ns}$ \\ \cline{2-7} 
         & Bob & $\qty{2}{\uW}$, $\qty{10}{\us}$ & $\num{0.13}P!s$, $\qty{80}{\us}$ & $\num{0.57}P!s$, $\qty{80}{\us}$ & $\num{0.8}P!s$, $\qty{12(1)}{\us}$ & $\num{50}P!s$, $\qty{2.3}{\ns}$ \\ \hline
        \multirow{2}{*}{\thead{Fig.~\ref{fig: supp - PLEs}(a) \\ (cr-checked data)}} & Alice & $\qty{1.5}{\uW}$, $\qty{10}{\us}$ & $\num{1.71}P!s$, $\qty{100}{\us}$ & \multirow{2}{*}{N/A} & \multirow{2}{*}{N/A} & \multirow{2}{*}{N/A} \\ \cline{2-4}
         & Bob & $\qty{2}{\uW}$, $\qty{10}{\us}$ & $\num{0.11}P!s$, $\qty{80}{\us}$ &  &  &  \\ \hline
        \multirow{2}{*}{Fig.~\ref{fig: supp - PLEs}(b)} & Alice & $\qty{1.5}{\uW}$, $\qty{10}{\us}$ & $\num{1.71}P!s$, $\qty{100}{\us}$ & \multirow{2}{*}{N/A} & \multirow{2}{*}{N/A} & \multirow{2}{*}{N/A} \\ \cline{2-4}
         & Bob & $\qty{2}{\uW}$, $\qty{10}{\us}$ & $\num{0.11}P!s$, $\qty{80}{\us}$ &  &  &  \\ \hline
        Fig.~\ref{fig: supp - optical pulses}(a) & Bob & $\qty{20}{\uW}$, $\qty{100}{\us}$ & $\num{0.16}P!s$, $\qty{80}{\us}$ & $\num{0.57}P!s$, $\qty{80}{\us}$ & N/A & varied, $\qty{2.3}{\ns}$ \\ \hline
        Fig.~\ref{fig: supp - optical pulses}(b) & Bob & $\qty{20}{\uW}$, $\qty{100}{\us}$ & $\num{0.18}P!s$, $\qty{80}{\us}$ & $\num{0.15}P!s$, $\qty{80}{\us}$ & N/A & varied \\ \hline
        Fig.~\ref{fig: supp - optical pulses}(c) & Bob & $\qty{1}{\uW}$, $\qty{10}{\us}$ & $\num{0.17}P!s$, $\qty{80}{\us}$ & $\num{0.61}P!s$, $\qty{80}{\us}$ & N/A & $\num{40}P!s$, $\qty{2.3}{\ns}$ \\ \hline
        \multirow{2}{*}{Fig.~\ref{fig: supp - Ramsey}} & Alice & $\qty{5}{\uW}$, $\qty{50}{\us}$ & $\num{0.17}P!s$, $\qty{150}{\us}$ & $\num{0.50}P!s$, $\qty{200}{\us}$ & $\num{0.67}P!s$, $\qty{15}{\us}$ & \multirow{2}{*}{N/A} \\ \cline{2-6}
         & Bob & $\qty{2}{\uW}$, $\qty{10}{\us}$ & $\num{0.10}P!s$, $\qty{80}{\us}$ & $\num{0.41}P!s$, $\qty{80}{\us}$ & $\num{0.55}P!s$, $\qty{16}{\us}$ &  \\ \hline
        Fig.~\ref{fig: supp - ODMR} & Bob & $\qty{2}{\uW}$, $\qty{10}{\us}$ & $\num{0.10}P!s$, $\qty{80}{\us}$ & $\num{0.41}P!s$, $\qty{80}{\us}$ & $\num{0.55}P!s$, $\qty{16}{\us}$ & N/A \\ \hline
        \multirow{2}{*}{Fig.~\ref{fig: supp - TPQI runs}(a)} & Alice & $\qty{1}{\uW}$, $\qty{50}{\us}$ & $\num{0.31}P!s$, $\qty{100}{\us}$ & $\num{1.23}P!s$, $\qty{200}{\us}$ & \multirow{2}{*}{N/A} & $\num{460}P!s$, $\qty{1.34}{\ns}$ \\ \cline{2-5} \cline{7-7} 
         & Bob & $\qty{5}{\uW}$, $\qty{100}{\us}$ & $\num{0.27}P!s$, $\qty{100}{\us}$ & $\num{5.49}P!s$, $\qty{200}{\us}$ &  & $\num{446}P!s$, $\qty{1.46}{\ns}$ \\ \hline
        \multirow{2}{*}{Fig.~\ref{fig: supp - TPQI runs}(b)} & Alice & $\qty{15}{\uW}$, $\qty{50}{\us}$ & $\num{0.11}P!s$, $\qty{150}{\us}$ & $\num{0.64}P!s$, $\qty{200}{\us}$ & \multirow{2}{*}{N/A} & $\num{4}P!s$, $\qty{1.7}{\ns}$ \\ \cline{2-5} \cline{7-7} 
         & Charlie & $\qty{10}{\uW}$, $\qty{100}{\us}$ & $\num{0.07}P!s$, $\qty{150}{\us}$ & $\num{0.32}P!s$, $\qty{200}{\us}$ &  & $\num{42}P!s$, $\qty{1.7}{\ns}$ \\ \hline
    \end{tabular}
\end{table*}
\renewcommand{\arraystretch}{1}

This section presents the experimental settings that were used to obtain the results from the main text. The optical powers and durations of all control pulses from the main text are summarized in Table~\ref{tab: optical settings}. Here, resonant optical powers are expressed as a fraction of saturation power in order to remove setup specific dependencies such as optical alignment and fiber-waveguide coupling efficiencies.

The settings for the CR checks are chosen to maximize the stability of the emitters as much as possible. The power of the optical pulse is calibrated throughout the experiments, which is why its power varies slightly between blocks of experiments. The same also holds true for the readout duration. The power of the optical pulse is estimated from the EOM suppression ratio, but is also influenced by the relative timing between the EOM and AOM pulses.

\section{Statistics of Emitter Characterization} \label{app: statistics}
As explained in Appendix~\ref{app: magnetic field}, the magnetic field magnitude and direction are chosen according to a strict set of constraints. Most notably, the maximum qubit splitting at which we can get sufficient microwave driving sets a limit of $\approx \qty{1}{\GHz}$ on the feasible tuning range of the optical transitions of a given emitter. This is more than an order of magnitude lower than the expected inhomogeneous distribution of resonance frequencies of SnV centers in the device \cite{brevoordLargerangeTuning2025}. This section gives an overview of the methodology that was used to find emitters that were suitable for the experiment.
\begin{figure*}
    \centering
    \includegraphics{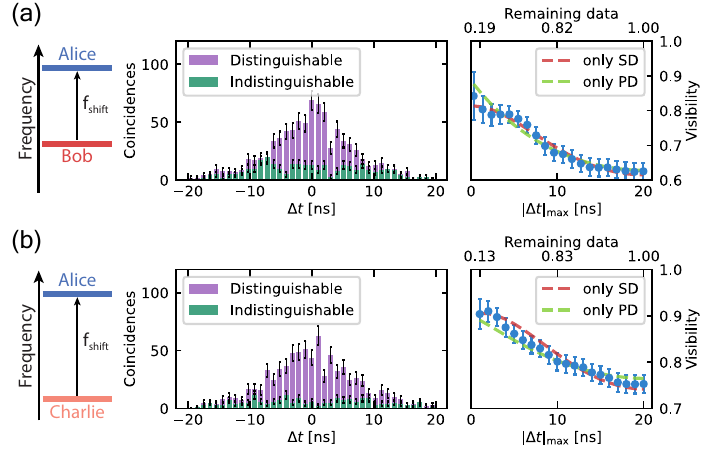}
    \caption{\textbf{Two-photon interference using photon frequency shifting.} \textbf{(a)} Two-photon interference experiment with the same pair of emitters as the results from the main text. Left: Schematic indicating the frequency shift of photons from the Bob emitter. Middle: Histogram of the coincidences as a function of measured photon detection time difference $\Delta t$. Right: Relation between the interference visibility and the maximum allowed detection time difference ($|\Delta t|!{max}$. \textbf{(b)} Same as in (a), but now with a different emitter (named Charlie) in a different waveguide on the Bob side.}
    \label{fig: supp - TPQI runs}
\end{figure*}

As visible in Fig.~\ref{fig:device}(b), the sample contains many waveguides that are optically accessible with the tapered fibers. After coupling the tapered fiber to a waveguide, a large-range PLE measurement is done over a \qty{30}{\GHz} frequency range. Such a measurement typically reveals a plethora of peaks, from which a selection of frequencies is made to investigate. At each frequency, we follow a fixed characterization protocol, in which we only proceed with characterization measurements for a given emitter if preceding characterization measurements on the emitter met our predefined requirements.

The investigation of an emitter starts by measuring optical properties at zero-field, which are mainly measured through the statistics of the CR checks and their stability. If stable CR checks are possible, a saturation curve measurement is done to reveal the collection efficiency of the specific emitter. We only continue characterization if the saturation power is below \qty{10}{\nW} and the collection efficiency in the PSB is above 1\%.

After these optical characterization measurements at zero-field, we turn on a \qty{100}{mT} magnetic field and sweep the direction of it over the 4 possible orientations of the emitters in the $\langle 110 \rangle$-oriented waveguides. Through PLE measurements we determine the orientation of the emitter's symmetry axis and proceed with spin control measurements at a \qty{100}{mT} aligned magnetic field. We start by calibrating the initialization and readout durations, after which an ODMR measurement is done to determine the qubit splitting. If the ODMR dip is found, a Rabi measurement is done to characterize the Rabi frequency, which we require to be above \qty{1}{\MHz}. Finally, a Hahn-Echo experiment is done to characterize the decoherence time of the emitter, which we require to be at least \qty{50}{\us}.

Lastly, the magnetic field is again turned off and a narrow CR-checked PLE scan (similar to Fig.~\ref{fig: supp - PLEs}(a)) is done to find the effective linewidth of the emitter under operating conditions similar to the entanglement experiment. Linewidths up to \qty{50}{\MHz} are accepted during the characterization.

In total, 90 emitters in 52 waveguides were characterized in this way. From this investigation, 4 pairs (Alice \& Bob) of candidate emitters are found that meet the above criteria, while having optical frequencies within \qty{5}{\GHz} from one another. The pair of emitters that is used for the experiment is chosen as it has the smallest frequency difference of all these pairs. 

\section{Frequency-Shifted Photon Interference} \label{app:AddTPQI}
As mentioned in the main text, photon interference can both be achieved by overlapping the optical transition frequencies of the emitters, or by shifting the frequency of their emitted photons. In this section, we present photon interference data for two additional experimental runs. The visibility dependence on photon detection time difference for both experiments can be found in Fig.~\ref{fig: supp - TPQI runs}. In both experiments, the two emitters have detuned optical transitions, and so the emitted photons from one of the emitters is shifted in frequency to overcome this detuning.

The data from Fig.~\ref{fig: supp - TPQI runs}(a) is obtained from the same pair of emitters as those used in the main text. The experiment is conducted at a detuning of $\qty{1.45}{\GHz}$. This detuning is achieved by exciting the set of spin-conserving transitions with the farthest splitting on the two emitters, while also using a different magnetic field from the experiments in the main text. We also show data from a separate experiment conducted with a different emitter on the Bob side (in a different waveguide, named Charlie in the schematic). For this pair, photon frequency shifting was necessary to overcome the \qty{3.16}{\GHz} detuning between the optical transition frequencies of the two emitters. 
\begin{figure*}
    \centering
    \includegraphics{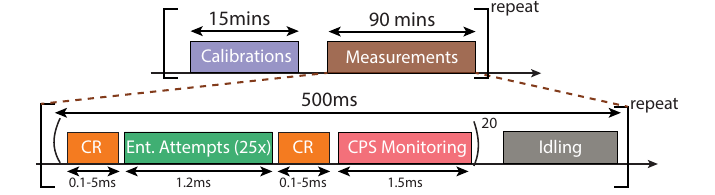}
    \caption{\textbf{Detailed breakdown of the entanglement sequence.} Top: 90 minute blocks of measurements are interleaved with 15 minute blocks of calibrations. Bottom: Blocks of measurements consist of CR checks, entanglement attempts, counts per shot (CPS) monitoring and idling. The combination of these ensure a $\qty{1}{\kHz}$ attempt rate.}
    \label{fig: supp - MVPE sequence}
\end{figure*}

Both datasets in Fig.~\ref{fig: supp - TPQI runs} show qualitatively similar behavior to the non-shifted result from Fig.~\ref{fig:tpqi}, thereby confirming the working principle of this implementation of frequency-shifted photon interference. However, visibilities are overall lower than in the non-shifted case due to the additional optical elements on Bob's side: first, there is significantly lower signal (and hence signal-to-noise) due to additional losses in the frequency-shifting path (Appendix~\ref{app:setup}), and second, the reflected part of the short excitation pulse is broadened in the time domain by the narrow filter cavity, leading to leakage of laser photons into the photon detection window.

\section{Details of the Entanglement Experiment}
\label{app:entanglement}
This section presents additional details of the entanglement experiments as presented in Fig.~\ref{fig:entanglement} of the main text.

\subsection{Entanglement Sequence \& Experimental Timings}
The sequence of the experiment is illustrated in Fig.~\ref{fig: supp - MVPE sequence}. The data for Fig.~\ref{fig:entanglement} is gathered over a week of continuous automatized runtime. Data is gathered in blocks of $\qty{1.5}{\hour}$, which are preceded by a $\qty{15}{\min}$ block of calibrations on both emitters. 

During measurements, we conservatively limit the attempt rate to $\qty{1}{\kHz}$ to avoid microwave-induced heating effects. The duration of a single entanglement attempt is much shorter than $\qty{1}{\ms}$ and we make use of the remaining idling time to interleave live monitoring of the CPS for stability analysis purposes (this feature is not further exploited in the current work but could be useful for more future automation).

After 20 blocks of 25 entanglement attempts and CPS monitoring, we record the total duration and ensure the $\qty{1}{\kHz}$ attempt rate by introducing an idle time for the leftover duration. The $\qty{1}{\kHz}$ rate is chosen according to a calibration of the spin coherence on both emitters in $\qty{1.5}{\hour}$ blocks at different attempt rates.

\subsection{Calibration and Heralding Parameters}
We define the parameter ranges that yield valid entanglement heralding events prior to the experiment based on independently measured parameters: we define the $\qty{20}{\ns} \approx 3\tau_\text{A/B}$ window in which we herald photon detection events and set the maximum detection time difference of valid heralded events to $\left|\Delta t\right|!{max} = \qty{10}{\ns}$ (grey dotted line in Fig.~\ref{fig:tpqi}(d) from the main text). Both of these filter metrics are available at the time of the herald and could validate the heralded entanglement in real-time - here we choose to also record data slightly outside these parameter ranges for analysis purposes. Similarly, metrics are defined beforehand for the calibration routines - not passing these metrics invalidates the following block of data taking. Future implementations can be more time-efficient by running automated calibration loops until successfully passing the threshold.

\subsection{Details on the Entanglement Attempts}
Each block of entanglement attempts is preceded by a CR check, which we operate in a regime of $\sim$$5-10\,\%$ success probability per probe pulse. We run these CR checks independently on both emitters and invalidate the success on one emitter after 10 failed checks on the other emitter.

The entanglement attempts themselves start with $\qty{100}{\us}$ ($\qty{80}{\us}$) of spinpumping for the Alice (Bob) emitter to initialize its spin state with an estimated infidelity of 0.002 (0.01). Next, we prepare an equal superposition state on both emitters through a $\frac{\pi}{2}$ rotation of the spin states about their rotating frame's x-axes. Since this initial $\frac{\pi}{2}$-pulse needs to be played regardless of measured photons, it is the dominant source of heating. After this, a $\qty{1.6}{\ns}$ ($\qty{2.3}{\ns}$) optical pulse resonantly excites the $\ket{0}$ part of the wavefunction, thereby creating an excitation in the "early" optical mode. We only want to proceed with the entanglement attempt in instances where this early photon is detected within the time window. The presence or absence of a detection signal from the SNSPDs is analyzed in real-time by an FPGA, which conditions the proceeding microwave $\pi$-pulse on the early heralding signal. In this way, we only play this second (heat-inducing) microwave pulse if a valid early photon is detected. This $\pi$-pulse is played exactly $\tau=\qty{8.62}{\us}$ after the initial $\frac{\pi}{2}$-pulse. Finally, a second optical pulse excites again the $\ket{0}$ part of the wavefunctions of both emitters and causes the generation of photons in the "late" optical mode. The time difference between the early and late time bin in our experiment is $\qty{2}{\us}$. Again, the detection signal is analyzed in real-time by an FPGA, which upon measurement of a second heralding signal flags the system to start with the tomography. The click pattern of the detectors directly determines if the $\ket{\Psi^{+}}$ (twice the same detector) or $\ket{\Psi^-}$ (two different detectors) Bell state is generated. Since this information is available at the time of the late herald signal, we extract it in the FPGA in order to play a virtual $Z-$ gate on the Bob emitter if we got clicks in the two time bins from the same detector. This means that we deterministically create the $\ket{\Psi^-}$ Bell state, on which we subsequently perform single-shot readout, where the desired basis is set by an appropriate microwave pulse. The temporal spacing between the tomography microwave pulse and the $\pi$-pulse between the time bins is again chosen to be exactly $\tau$ in order to filter out any quasi-static noise. 

\subsection{Outlook on Improving Entanglement Success Probability}
\label{app: ideal entanglement prob estimate}
A significant increase in entanglement success probability can be obtained by switching to an entanglement scheme based on photonic Fock-state encoding \cite{hermansEntanglingRemoteQubits2023}. In such a scheme, the success probability is given by 
\begin{equation} \label{eq: single click success prob}
    P_\text{success} = 2\alpha\eta,
\end{equation}
where $\alpha$ is the experimental setting of the 'bright'-state population in the prepared imbalanced superposition (below we use $\alpha=0.05$) and $\eta$ is the probability of detecting a coherent ZPL photon after resonant excitation. Here, we estimate a value of $\eta$ that could be achieved in the current setup if the near-term improvements mentioned in the main text are implemented (in particular the reflector in the waveguide and in-contact fiber coupling). We subsequently use that estimate of $\eta$ to derive feasible entanglement generation rates.

Firstly, the inherent coupling from the SnV center's dipole moment (with a relevant ZPL fraction of 0.36) to the waveguide mode is estimated from simulations to be around 0.6. With an in-contact coupled tapered optical fiber, coupling efficiencies of the waveguide-fiber interface exceeding 0.9 have been reported \cite{burekFiberCoupledDiamond2017}. Insertion loss of the etalon cavity that filters out the incoherent D-transition, inherent losses from the cross-polarization scheme and additional losses are estimated to add up to a transmission of 0.6. The time filtering of the laser reflections from the short optical pulse leads to another factor of 0.9. Finally, the photon detectors could operate at a 0.95 efficiency. The product of all these factors leads to an estimate of $\eta$ of 0.1, yielding an estimated entanglement success probability of $P_\text{success} = 0.01$.

In the longer term, as discussed in the outlook, photonic integration may enable the parallel operation on thousands of qubits per node. A ballpark estimate for the resulting entangling rate at a node distance of \qty{50}{\km} can be estimated as follows. The roundtrip communication time to the heralding station (midway at \qty{25}{\km}) is \qty{250}{\us}, setting the entanglement attempt rate $r_\text{attempt}$ per qubit at \qty{4}{\kHz}. We then consider an architecture with an overall multiplexing factor $N_\text{multiplex}$ of 2500 (e.g. combining 64x time-multiplexing and 40x frequency-multiplexing). We include additional photon loss of \qty{20}{\dB} accounting for photon routing, quantum frequency conversion and attenuation in \qty{25}{\km} of optical fiber. The success probability per attempt per qubit is then $P_\text{success} = 0.0001$, yielding a system-level entangling rate $P_\text{success} \cdot N_\text{multiplex} \cdot r_\text{attempt}$ of about \qty{1}{\kHz}.

\section{Overview of fit functions and parameters}
Table~\ref{tab: fits and parameters} provides an overview of the fit models that were used to analyze the data in the figures of this work. The table additionally provides quantitative values for the key metrics that are obtained from the fits.

\renewcommand{\arraystretch}{1.3}
\begin{table*}
    \centering
    \caption{\textbf{Fit function and parameter values for all figures.} Overview of the used fit functions and the extracted parameters from the fits in all figures of this work. Errors for the parameter values are extracted from the fit uncertainties.}
    \label{tab: fits and parameters}
    \begin{tabular}{|c|c|c|cc|}
        \hline
        Figure & Fit function & Parameter name & \multicolumn{1}{c|}{Emitter} & Parameter value \\ \hline
        \multirow{4}{*}{Fig. 2(a)} & \multirow{4}{*}{$A \text{e}^{- t/\tau}$} & \multirow{2}{*}{Lifetime ($\tau$)} & \multicolumn{1}{c|}{Alice} & $\qty{7.63(4)}{\ns}$ \\ \cline{4-5} 
         &  &  & \multicolumn{1}{c|}{Bob} & $\qty{7.11(3)}{\ns}$ \\ \cline{3-5} 
         &  & \multirow{2}{*}{ZPL detection probability ($\eta$)} & \multicolumn{1}{c|}{Alice} & $\qty{0.1790(6)}{\%}$ \\ \cline{4-5} 
         &  &  & \multicolumn{1}{c|}{Bob} & $\qty{0.1946(6)}{\%}$ \\ \hline
        \multirow{2}{*}{Fig. 2(b)} & \multirow{2}{*}{$A \cos(\Omega t)$} & \multirow{2}{*}{Rabi frequency ($\Omega$)} & \multicolumn{1}{c|}{Alice} & $\qty{1.194(2)}{\MHz}$ \\ \cline{4-5} 
         &  &  & \multicolumn{1}{c|}{Bob} & $\qty{2.187(3)}{\MHz}$ \\ \hline
        \multirow{3}{*}{\thead{Fig. 2(c) \\ (left panel)}} & \multirow{3}{*}{$A e^{-(x/T_2)^n} + 0.5$} & Decoherence time ($T_2$) & \multicolumn{2}{c|}{See right panel} \\ \cline{3-5} 
         &  & \multirow{2}{*}{Exponential scaling} & \multicolumn{1}{c|}{Alice} & $1.2(2)\text{--}3.0(7)$ \\ \cline{4-5} 
         &  &  & \multicolumn{1}{c|}{Bob} & $2.8(6)\text{--}3.0(9)$ \\ \hline
        \multirow{2}{*}{\thead{Fig. 2(c) \\ (right panel)}} & \multirow{2}{*}{$A  T_2^\chi$} & \multirow{2}{*}{Scaling ($\chi$)} & \multicolumn{1}{c|}{Alice} & $0.71(7)$ \\ \cline{4-5} 
         &  &  & \multicolumn{1}{c|}{Bob} & $0.84(4)$ \\ \hline
        \multirow{2}{*}{Fig. 3} & \multirow{2}{*}{See Appendix \ref{app:model_tpqi}} & (Only) Spectral diffusion & \multicolumn{1}{c|}{\multirow{2}{*}{Both}} & $\qty{34.2(2)}{\MHz}$ \\ \cline{3-3} \cline{5-5} 
         &  & (Only) Pure dephasing & \multicolumn{1}{c|}{} & $\qty{27(1)}{\MHz}$ \\ \hline
        \multirow{2}{*}{Fig. 6(a)} & \multirow{2}{*}{$\text{Voigt}(f, \gamma, \sigma)$} & \multirow{2}{*}{Full-Width at Half Maximum} & \multicolumn{1}{c|}{Alice} & $\qty{23.7(7)}{\MHz}$ \\ \cline{4-5} 
         &  &  & \multicolumn{1}{c|}{Bob} & $\qty{29.0(4)}{\MHz}$ \\ \hline
        \multirow{2}{*}{\thead{Fig. 6(b) \\ (right panel)}} & \multirow{2}{*}{$\Gamma_0 \sqrt{1 + s}$} & \multirow{2}{*}{Linewidth ($\Gamma_0$)} & \multicolumn{1}{c|}{Alice} & $\qty{27(4)}{\MHz}$ \\ \cline{4-5} 
         &  &  & \multicolumn{1}{c|}{Bob} & $\qty{30(1)}{\MHz}$ \\ \hline
        \multirow{3}{*}{Fig. 7(a)} & \multirow{3}{*}{$A\left(1 - \text{e}^{-\beta\sqrt{P}} \right) + A \text{e}^{-\beta\sqrt{P}} \sin^2\left(\frac{\alpha\sqrt{P}}{2}\right)$} & $\pi$-pulse power & \multicolumn{1}{c|}{Bob} & $\qty{43.1(7)}{P_s}$ \\ \cline{3-5} 
         &  & Experimental power ($0.85\pi$) & \multicolumn{1}{c|}{Bob} & $\qty{31.2(5)}{P_s}$ \\ \cline{3-5} 
         &  & PSB detection probability & \multicolumn{1}{c|}{Bob} & $\qty{0.746(9)}{\%}$ \\ \hline
        \multirow{2}{*}{Fig. 8} & \multirow{2}{*}{$A \text{e}^{(-t/{T_2}^{*})^2} \sin(\Delta t + \phi) + B$} & \multirow{2}{*}{Dephasing time ($T_2^{*}$)} & \multicolumn{1}{c|}{Alice} & $\qty{2.32(7)}{\us}$ \\ \cline{4-5} 
         &  &  & \multicolumn{1}{c|}{Bob} & $\qty{1.22(3)}{\us}$ \\ \hline
        \multirow{2}{*}{Fig. 9} & \multirow{2}{*}{$1 - \sum\limits_{i=1}^4 \text{Lorentzian}(f_i, \gamma_i) $} & Splitting ($f_4 - f_1$) & \multicolumn{1}{c|}{Bob} & $\qty{2.42(2)}{\MHz}$ \\ \cline{3-5} 
         &  & Center frequency ($\frac{f_2 + f_3}{2}$) & \multicolumn{1}{c|}{Bob} & $\qty{2.32571(5)}{\GHz}$ \\ \hline
        \multirow{2}{*}{\thead{Fig. 10 \\ (Exp. 1)}} & \multirow{2}{*}{See Appendix \ref{app:model_tpqi}} & (Only) Spectral diffusion & \multicolumn{1}{c|}{\multirow{2}{*}{Both}} & $\qty{24.9(4)}{\MHz}$ \\ \cline{3-3} \cline{5-5} 
         &  & (Only) Pure dephasing & \multicolumn{1}{c|}{} & $\qty{15(1)}{\MHz}$ \\ \hline
        \multirow{2}{*}{\thead{Fig. 10 \\ (Exp.1)}} & \multirow{2}{*}{See Appendix \ref{app:model_tpqi}} & (Only) Spectral diffusion & \multicolumn{1}{c|}{\multirow{2}{*}{Both}} & $\qty{40.3(6)}{\MHz}$ \\ \cline{3-3} \cline{5-5} 
         &  & (Only) Pure dephasing & \multicolumn{1}{c|}{} & $\qty{37(1)}{\MHz}$ \\ \hline
    \end{tabular}
\end{table*}
\renewcommand{\arraystretch}{1}

\section{Model TPQI}
\label{app:model_tpqi}

By investigating the photon arrival time difference ($\Delta t$) dependence of the visibility, we can gain information on the combined spectral diffusion and pure dephasing of the involved emitters. To this end, we model the photon probability function to be composed of two modes: an ideal single photon emitted by the tin-vacancy center $p!{s}$ and some background noise source $p!{n}$, e.g. laser photons or photons of other emitters. The probability of a photon arriving from emitter $A$ at the central beamsplitter at time $t$ is hence simply

\begin{equation}
    p!A(t) = p!{s}^A(t) + p!{n}^A(t).
\end{equation}

The probability distribution for the single photon can be written as 
\begin{equation}
    \label{eq:prob_photon}
    p!{s}^A(t) = N!A e^{-\frac{t}{\tau!A}} \; \Theta(t)\Theta(T!{det}-t),
\end{equation}
with $N!A = \frac{\eta!A}{\tau!A\left(1-e^{-T!{det}/\tau!A}\right)}$ as a normalization factor to ensure the total probability of detecting a photon in the detection window $T!{det}$ is $\eta!A$, $\tau!A$ being the emitter lifetime, and $\Theta$ being the heaviside function.

\subsection{Distinguishable scenario}
In the distinguishable setting of a TPQI experiment, no interference occurs between the photons. We can hence express the probability of a coincidence in the output ports $C$ and $D$ of the beamsplitter with reflectivity and transmittivity parameters $R$ and $T$ as

\begin{align}
\begin{split}
    p!{coinc}^\text{dist}(t, \Delta t) &= g!{CD}(t, t+\Delta t) \\
    &= \sum_{i,j,L,M} K_{ij} g_{i \;j}^{L \; M}(t, t+\Delta t),
\end{split}
\end{align}

where $t$ is the detection time of the first photon, $K_{ij} \in \{R^2, RT, T^2\}$ is a prefactor to select the right paths for modes $i$ and $j$, and $g_{ij}(t, t+\Delta t)$ describes the correlation function between photon mode $i$ and mode $j$
\begin{equation}
    \label{eq:correlation_photons}
    g_{i \;j}^{L \; M}(t, t+\Delta t) = \langle p_{i}^L(t)p_{j}^M(t+\Delta t)\rangle.
\end{equation}

As an example, $g!{s \; s}^{A \; B}(t, t+\Delta t)$ is the correlation function for two ideal single photons from emitters $A$ and $B$ arriving at detector $C$ and $D$ at time $t$ and $(t+\Delta t)$, respectively. The correct prefactor for this term is $T^2$ as both photons need to be transmitted through the beamsplitter for this scenario. Importantly, the auto-correlation $g!{s \; s}^{A \; A}$ and $g!{s \; s}^{B \; B}$ vanish as we model the photons as perfect single photons.

Assembling all non-vanishing correlation functions, and integrating over all possible detection times $t$ within the detection window, we find

\begin{align}
\label{eq:p_coinc_dist}
\begin{split}    
p!{coinc}^\text{dist}(\Delta t) 
= T^2 &\left(g!{s \; s}^{A \; B}(\Delta t) + g!{s \; n}^{A \; B}(\Delta t) + g!{n \; s}^{A \; B}(\Delta t) + g!{n \; n}^{A \; B}(\Delta t)\right) \\
+ R^2 &\left(g!{s \; s}^{B \; A}(\Delta t) + g!{s \; n}^{B \; A}(\Delta t) + g!{n \; s}^{B \; A}(\Delta t) + g!{n \; n}^{B \; A}(\Delta t)\right) \\
+ RT &\left(g!{s \; n}^{A \; A}(\Delta t) + g!{n \; s}^{A \; A}(\Delta t) + g!{n \; n}^{A \; A}(\Delta t)\right) \\ 
+ RT &\left(g!{s \; n}^{B \; B}(\Delta t) + g!{n \; s}^{B \; B}(\Delta t) + g!{n \; n}^{B\; B}(\Delta t)\right).
\end{split}
\end{align}

The correlation functions can be calculated independently. In the distinguishable scenario, $g!{s \; s}^{A \; B}(\Delta t)$ can be easily determined by inserting the individual exponential probability distributions of the ideal single photons from Eq.~\ref{eq:prob_photon} into Eq.~\ref{eq:correlation_photons} to

\begin{equation}
    g!{s \; s}^{A \; B}(\Delta t) = \frac{N!AN!B}{\alpha e^{\frac{\Delta t}{\tau!B}}} \left(e^{-\alpha a} - e^{-\alpha b}\right) \; \Theta\left(T!{det}-\left|\Delta t\right|\right),
\end{equation}
where $\alpha = \left(\tau!A^{-1} + \tau!B^{-1}\right)$, $a = \max\left(0, -\Delta t\right)$ and $b = \min\left(T!{det}, T!{det}-\Delta t\right)$. 

For all terms involving noise contributions, an assumption about the noise has to be made to calculate its correlation with the signal photons as well as its autocorrelation. For the sake of simplicity, we limit ourselves to modeling the noise contribution as single photons (e.g. from another emitter close in transition frequency). This leads to the same expression for $g!{s \; n}^{A \; B}(\Delta t)$ as for $g!{s \; s}^{A \; B}(\Delta t)$ in the distinguishable case and vanishing auto-correlations. A more general treatment would incorporate the precise noise contributions and their exact photon statistics. For the modeling here, we will apply only single photon noise models.

\subsection{Indistinguishable Scenario}
\begin{figure*}
	\centering
	\includegraphics[width=\linewidth]{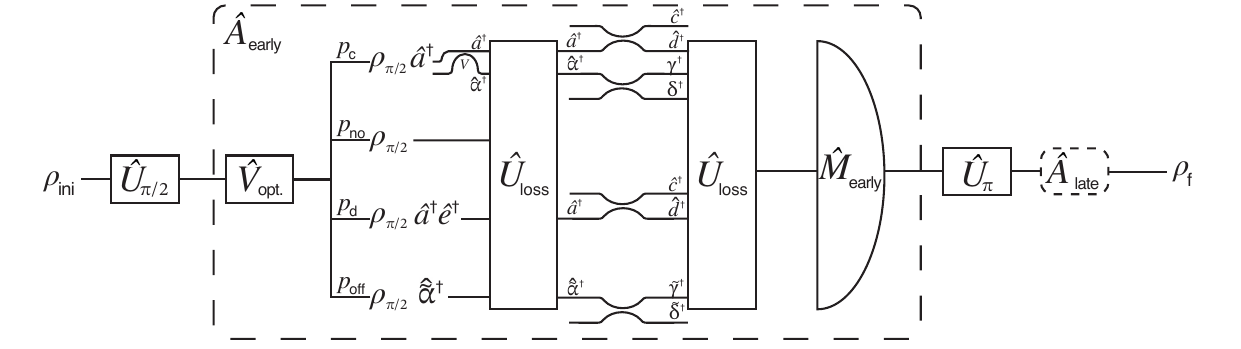}
	\caption{\textbf{Error model for the Barrett-Kok protocol.} The different steps performed on a single qubit during the protocol as well as the photonic modes involved are presented. A detailed description of all steps can be found in the main text.}
	\label{fig:error-flow}
\end{figure*}
In the indistinguishable experiment setting, the single photons from $A$ and $B$ interfere, resulting in $g!{s \; s}^{A \; B} = 0$. However, an imperfect beamsplitter, different emitter lifetimes, spectral diffusion and pure dephasing lead to non-vanishing correlations. The correlation function is derived in Ref. \cite{kambsLimitationsIndistinguishability2018}. We add a limited detection time window to this model, which results in 

\begin{align}
    g!{s \; s}^{A \; B}(\Delta t) = \mathcal{G}^{(2)}!{0}(\Delta t) + \mathcal{G}^{(2)}!{int}(\Delta t),
\end{align}
with
\begin{align}
    \mathcal{G}^{(2)}!{0}(\Delta t) &= T^2 g!{s \; s}^{A \; B}(\Delta t) + R^2 g!{s \; s}^{B \; A}(\Delta t)\\
    \begin{split}
    \mathcal{G}^{(2)}!{int}(\Delta t) &= \frac{-2RT e^{-\gamma\left|\Delta t\right| - 2\pi^2\Sigma^2\Delta t^2}}{\left(\tau!A + \tau!B\right) \left(1-e^{-\frac{T!{det}}{\tau!A}}\right)\left(1-e^{-\frac{T!{det}}{\tau!B}}\right)} \cdot\\
    &\left( 1 - e^{-\alpha\left(T!{det}-\left|\Delta t\right|\right)} \right) \cdot
    \begin{cases}
        1 &\text{if} \; 
        \left|\Delta t\right| \leq T!{det}, \\
        0 &\text{else}.
    \end{cases}
    \end{split}
\end{align}

Here, we define 
\begin{align}
    \gamma &= \Gamma!A^* + \frac{1}{2\tau!A} + \Gamma!B^* + \frac{1}{2\tau!B}, \\
    \Sigma^2 &= \sigma!A^2 + \sigma!B^2,
\end{align}
with $\Gamma!{A/B}^*$ as the pure dephasing rate (in natural frequency) and $\sigma!{A/B}$ the standard deviation of the Gaussian spectral diffusion profile (in natural frequency). From these definitions, it is clear that a measurement of $\mathcal{G}^{(2)}!{int}(\Delta t)$ only allows the extraction of the combined pure dephasing and spectral diffusion magnitudes, not the individual ones. These values are the spectral diffusion and pure dephasing magnitudes as quoted in Table \ref{tab: fits and parameters}.

With this modified expression for $g!{s \; s}^{A \; B}$, $p!{coinc}^\text{indist}(\Delta t)$ takes the same form as Eq.~\ref{eq:p_coinc_dist}. We then find the visibility for a maximum allowed detection time difference $\Delta t!{max}$ as

\begin{equation}
    \label{eq:visibility_dt}
    V(\Delta t!{max}) = 1-\frac{\int_{-\Delta t!{max}}^{\Delta t!{max}}p!{coinc}^\text{indist}(\Delta t) d\Delta t}{\int_{-\Delta t!{max}}^{\Delta t!{max}}p!{coinc}^\text{dist}(\Delta t) d\Delta t}.
\end{equation}

\section{Model Entanglement}
\label{app:model_entanglement}
\renewcommand{\arraystretch}{1.3}
\begin{table*}
    \centering
    \caption{\textbf{Estimated errors for the experiment.} The values are extracted from supporting calibration measurements. We assume negligible dark count contributions for the simulation. For each error source, we consider two scenarios: First, we show the fidelity of the entangled state if the source was the only imperfection in the simulation. Secondly, we show the fidelity if all other errors were present except the reported error. The base fidelity of the simulation with all error sources is $\mathcal{F}=0.763$.}
    \label{tab:entanglment-errors}
    \begin{tabular}{|c|c|c|c|}
        \hline
        Source of infidelity & Value for simulation & \thead{Expected fidelity if \\ only source present} & \thead{Expected fidelity if \\ source removed} \\ \hline
        $V$ & 0.861 & 0.930 & 0.811 \\ \hline
        $R!{bs}$ & 0.542 & 0.999 & 0.765 \\ \hline
        $T!{bs}$ & 0.458 & 0.999 & 0.765 \\ \hline
        $p!{ini}^A$ & 0.002 & 0.998 & 0.764 \\ \hline
        $p!{ini}^B$ & 0.010 & 0.990 & 0.769 \\ \hline
        $p!{\pi/2}^A$ & 0.020 & 0.990 & 0.769 \\ \hline
        $p!{\pi/2}^B$ & 0.010 & 0.995 & 0.766 \\ \hline
        $p!{\pi}^A$ & 0.020 & 0.980 & 0.776 \\ \hline
        $p!{\pi}^B$ & 0.010 & 0.990 & 0.769 \\ \hline
        $p!{double}^A$ & 0.053 & 0.974 & 0.781 \\ \hline
        $p!{double}^B$ & 0.092 & 0.954 & 0.795 \\ \hline
        $p!{off}^A$ & 0.008 & 0.977 & 0.780 \\ \hline
        $p!{off}^B$ & 0.019 & 0.945 & 0.804 \\ \hline
        $p!{no-exc}^A$ & 0.050 & 1.000 & 0.765 \\ \hline
        $p!{no-exc}^B$ & 0.050 & 1.000 & 0.766 \\ \hline
    \end{tabular}
\end{table*}
\renewcommand{\arraystretch}{1}

In real-life experiments, all of the many steps involved in the Barrett-Kok protocol \cite{barrettEfficientHighfidelity2005} are subject to errors. In this section, we present an analytical model treating these errors and derive an expression for the success probability and fidelity. We select the error sources that are expected to be the most prominent in our system. The flow of the protocol for one of the qubits in a schematic circuit-diagram is depicted in Fig.~\ref{fig:error-flow}. The qubit start out in a mixed state $\rho!m$ and is prepared into an eigenstate $\rho!{ini}=\ketbra{0}{0}$ with error probability $p!{ini}$ to actually create the orthogonal eigenstate $\ketbra{1}{1}$. Afterwards, applying a $\pi/2$-gate with error probability $p_{\pi/2}$ creates an equal superposition state. 

Then, an optical pulse excites the qubit and possibly creates a photon. We consider four scenarios: In the error-free case with probability $p!c$, a coherent single photon $\hat{a}^\dagger$ is created from the $\ket{0}$-part of the superposition. To account for finite Hong-Ou-Mandel (HOM) visibility of the photons, it is split into two modes, where the indistinguishable mode $\hat{a}^\dagger$ will later interfere with mode $\hat{b}^\dagger$ from the other qubit and the distinguishable mode $\hat{\alpha}^\dagger$ does not interfere with the photon from the other qubit. Secondly, with probability $p!{no-exc}$, the superposition state is not excited (as we do not necessarily apply a full $\pi$-rotation on the optical transition) and is not connected to a photonic mode. The last two cases cover double excitation with a probability $p!{d}$, where a second mode $\hat{e}^\dagger$ is populated and later traced out; and off-resonant excitation with a probability $p!{off}$, where the photon in mode $\hat{\tilde{\alpha}}^\dagger$ stems from the $\ket{1}$-part of the superposition and is considered incoherent.

All photonic channels are then subject to loss before entering the HOM beamsplitter. Here, only the indistinguishable modes $\hat{a}^\dagger$ and $\hat{b}^\dagger$ from both qubits are interfered with each other, all distinguishable modes from both qubits get treated with separate beamsplitter modes. 
Another loss channel is followed by a measurement of the photons. This concludes the early timebin. After a $\pi$-rotation on the qubits, a second round of optical excitation, interference and detection takes place. Finally, the two-qubit state $\rho!f$ is obtained.

A detailed description of the error model can be found in Ref. \cite{waasTinVacancyCenter2026}.

Using this analytical model we calculate the expected correlations and fidelity of the heralded entanglement in Fig.~\ref{fig:entanglement}(c). In Table~\ref{tab:entanglment-errors} the extracted probabilities of the various error sources are shown. We show how big their effect would be if it was the only error source and what the entangled state fidelity would be if that error source was removed.

\newpage

\bibliography{bibliography}

\end{document}